\input epsf

\title{De Broglie geometry eliminating the infinities of QED;
An exact derivation of the Lamb shift formula in the normal case}

\newcommand{\cmt}[1]{\ifhmode\newline\fi{\sf *** \ \ #1 \\}}

\documentclass[11pt]{article}
\usepackage{latexsym}
\usepackage{epsfig}

\usepackage{amsmath}

\usepackage{amssymb}

\usepackage{exscale}

\usepackage{amsthm}
\usepackage{verbatim}

\def\:{\colon}

\long\def\onefigure#1#2{%  #1 picture,  #2  caption
\begin{figure*}[tbh]
\begin{center}
#1
\end{center}
\caption{#2}
\end{figure*}
} %end onefigure def

\newcommand{\iipefig}[1]  % immediate Ipe figure
{\smallskip\begin{center}\def\IPEfile{#1.ipe}\input{\IPEfile}\end{center}\smallskip}
\newcommand{\lipefig}[2]  % labeled Ipe figure
{\onefigure{\def\IPEfile{frh-#1.ipe}\input{\IPEfile}}{\label{f:#1} #2} }

\author
{Z. I. Szab\'o\thanks{Partially supported by
NSF grant DMS-0604861}
\thanks{Lehman College of CUNY, Bronx, NY 10468,USA, and 
R\'enyi Institute, Budapest,
POBox 127, 1364 Hungary. E-mail: zoltan.szabo@lehman.cuny.edu}}
\date{}
\begin{document}

\maketitle
%\rightline{Hatvanadik sz\"ulet\'esnapomra}
\begin{abstract}
This paper evolves a new non-perturbative theory by which the problem 
of infinities appearing in quantum physics can be handled. Its 
most important application is an exact derivation of the Lamb 
shift formula by using no renormalization. The 
Lamb shift experiment (1947) gave rise to one of the greatest
challenges whose explanation brought the modern 
renormalization technique into life. Since then this is the only tool for
handling these infinities. The relation between this  
renormalization theory and our non-perturbative theory is also 
discussed in this paper. 

Our key insight is the realization that the 
natural complex Heisenberg group representation splits the Hilbert space, 
$L^2_{\mathbb C}(\mathbb R^{2\kappa})$,
of complex valued
functions defined on an even dimensional Euclidean 
space into irreducible subspaces 
(alias zones) which are invariant also under the action 
of the Landau-Zeeman operator. 
After a natural modification, 
also the Coulomb operator can be involved into this zonal theory. 
Thus these zones can be  
separately investigated, both from 
geometrical and physical point of view.
In the literature only the zone 
spanned by the holomorphic polynomials has been investigated so far. 
This zone is the well known Fock space. 
This paper explicitly
explores also the ignored (infinitely many) other zones. 
It turns out that
quantities appearing as infinities on the total Hilbert space
are finite in the zonal setting. Even the zonal Feynman integrals
are well defined. In a sense,
the desired finite quantities are provided here by an extended 
particle theory where these extended objects show up also on the 
rigorously developed mathematical level. Name {\it de Broglie geometry} 
was chosen to suggest this feature of the zonal theory.
  
\end{abstract}
\section{Introduction.}
This paper consists of three parts. In the first two chapters all those
mathematical and physical structures are reviewed which are 
most essential to the third one entitled
``Interaction with the Coulomb field".

In the {\it first chapter}  
the  Zeeman-Hamilton operator, determined for a system of free charged 
particles orbiting in constant magnetic fields, is established as the 
Laplacian on a Riemannian, so called, Zeeman manifold. As far as the 
author knows, this exact matching of a physical Hamiltonian with the 
Laplacian of a Riemann manifold has never been recognized in the 
literature so far. Although the manifolds with such coincidence 
can be introduced in a most general way,  
in this paper only those defined by center periodic metric 2-step 
nilpotent Lie groups, $\Gamma_Z\backslash N$, are considered. In this 
formula the $\Gamma_Z =\{Z_\gamma\}$ is a partial lattice defined only 
on the center $\mathbf z$ of the metric group $(N,g)$, where $g$ is a 
left-invariant Riemann metric on $N$.
The factor manifold is then a trivial torus bundle, 
$\mathbf R^k\times T^l$, on which the metric 
can be briefly described as follows. Both on the base $\mathbf R^k$ 
and the torus $T^l=\Gamma_Z\backslash\mathbf z$, 
the induced metrics 
are flat (Euclidean), but, at a
point $(X,Z)$, the tangent spaces 
$T_X(\mathbf R^k)$
and
$T_Z(T^l)$
are not perpendicular. This property shows that the product $\times$ is 
just topological and not metric. 
In this paper mostly Heisenberg-type groups (see the definition later) 
will be considered. This particular groups have even dimensional 
X-spaces, 
$
\mathbf R^k=
\mathbf R^{2\kappa}
$, 
for which the dimension, $l$, of the Z-space can be 
arbitrary natural number. 

The natural $L^2$ Hilbert space defined on the torus bundle is 
subjugated, first, to a {\it primary splitting} 
$
L^2_{\mathbb C}(\mathbf R^k\times T^l)= 
\sum_\gamma W_\gamma
$.
Then, in the {\it secondary splitting}, each subspace 
$W_\gamma$ is further decomposed into the zones 
indicated in the Abstract.
The primary splitting is nothing but the Fourier-Weierstrass decomposition 
by means of the $\Gamma_Z$-periodic Fourier functions 
$e^{2\pi\langle Z_\gamma ,Z\rangle\mathbf i}$ 
defined by means of the lattice
points $Z_\gamma\in\Gamma_Z$ on the torus $T^l$. The above $L^2$ Hilbert 
space on the total space decomposes into orthogonal subspaces, 
$W_\gamma$, 
spanned by functions of the form 
$\Psi (X,Z)=\psi (X)e^{2\pi\langle Z_\gamma ,Z\rangle\mathbf i}$, 
where $\psi (X)$ is an $L^2$ function defined on $\mathbf R^k$.  

The Riemannian Laplacian, $\Delta$, on the total space is described in 
formula (\ref{Delta}).
This is not yet the Zeeman operator, however, its action 
on a function $\Psi$ 
can be written in the form  
$\Delta \Psi_\gamma =\Box_\gamma (\psi )(X)
e^{2\pi\langle Z_\gamma ,Z\rangle\mathbf i}$,
i. e., it leaves each $W_\gamma$ invariant and 
induces action only on function $\psi (X)$ depending just on $X$. 
This latter operator is explicitly described in formula (\ref{Box}). 
Comparison with (\ref{Zee_int}) reveals that operator
$-(1/2)\Box_\gamma$ is nothing but a Zeeman 
operator satisfying $V=0$. Although it is acting 
on functions depending just on the 
space-variables, the $\Box_\gamma$ is not a sub-Laplacian. In fact,  it
is obtained on the invariant subspaces defined on the 
total space and not by the submersion of the total space onto the
base $\mathbf R^k$. The 
2D version of this Hamiltonian was introduced by Landau,
in 1928. By our construction, it is obtained on a 3 
dimensional time-periodic Heisenberg group 
endowed with the natural left-invariant metric. In this paper, 
the explicit computations will be demonstrated
mostly on this 2D version. The general
theory is developed in \cite{sz6}. 

There is pointed out in Section\ref{interpret} that
this model is in strong relationship with Dirac's those relativistic
multi-time model, where the relativity is furnished by attributing
self-times to the particles.
(This theory differs from his relativistic electron theory.) 
Our model is attached to $\kappa =k/2$
particles orbiting in complex planes in constant magnetic fields. The
latter fields pin down unique inertia systems which define the self-times
of the particles. The time coordinates measured in this process appear
on the center of the group, however, the time is a secondary concept
in this theory. The primary objects are the angular momentum
endomorphisms defining the constant magnetic fields.

{\it The second theme} of the paper involves and further develops the 
Fock-Bargmann representation of the complex Heisenberg groups. 
From now on the investigations are performed on the X-space, 
$\mathbb R^k$. They are understandable without the mathematical 
model described above and it can be considered
as spectral analysis of the Zeeman operator. It is connected with
the above model such that the invariant subspaces $W_\gamma$ defined
by the primary decomposition are further investigated, which, by the
map $\Psi_\gamma (X,Z)\to \psi(X)$, can be identified with the function 
space 
$L^2_{\mathbf C}(X)$ 
defined just on the X-space. By the above discussions, 
the Laplacian $\Delta$ appears on a fixed invariant subspace as the
Landau-Zeeman operator $\Box_\lambda$ acting on this function space. 
The zones are defined by a spectral
decomposition of 
$L^2_{\mathbf C}(X)$,
thus it can be regarded as a secondary decomposition of the subspaces
obtained in the primary decomposition. 

This {\it secondary splitting} is defined by the Fock-Bargmann 
representation of the complex Heisenberg group. In
the literature, this representation is considered only on the Fock 
space, generated by the holomorphic polynomials in the total space of
complex valued functions defined on $\mathbf R^k$. No other invariant 
subspaces of this reducible representation have been investigated 
so far. This paper explores all irreducible subspaces,
called Zeeman zones, of the FB-representation. This zonal spectral 
analysis includes the 
explicit description of various zonal objects such as the projection 
kernels, the zonal spectra, and the zonal Wiener- resp. 
Schr\"odinger-flows. The most surprising result is that both zonal 
flows are of the trace class, defining 
the zonal partition and zeta functions in the standard way without
renormalization. Even the zonal Feynman measures on the 
path-space are well defined. In other words, quantities appearing 
as infinities on the global level are well defined finite ones 
in the zonal setting.

{\it The third and most important chapter} includes
the (so far ignored) Cou\-lomb operator $V$ in the investigations. 
Since it does not commute with the Zeeman-Landau operator, this operator
defines transmissions $V^{(a,b)}:\mathcal H^{(a)}\to\mathcal H^{(b)}$
between the zones. Such a map is defined by projecting the product
$V\varphi^{(a)}$, where $\varphi^{(a)}\in \mathcal H^{(a)}$, 
onto $\mathcal H^{(b)}$. Then $V^{(a,a)}$ maps 
$\mathcal H^{(a)}$ onto itself and is called zonal Coulomb operator.
In order to keep the zones invariant, only these zonal Coulomb operators
are retained. The complete zonal Zeeman operators are defined by 
$-(1/2)\Box^{(a)}-V^{(a,a)}$ and
the other transmission operators are omitted. The most remarkable 
features of the zonal Coulomb operators are that they commute with the
zonal Zeeman-Landau operator and their discrete spectrum appears on
eigenfunctions which are common with the Landau-Zeeman operator. 
Furthermore, they become integral
operators exhibiting local interactions. This phenomenon is to the
contrary of the global interaction
characteristic for the total operator $V$. However, both the trace and
$L^2$-norm of $V^{(a,a)}$ are infinities. This means that this operator
still defines 
infinite energy summations and is not yet suitable for analyzing
the Lamb shift. 

In a simple trace-computation the eigenvalues are summed up without 
any probabilistic distinction (equipartition principle) which often 
has caused problems and false conclusions in quantum physics.  
In many cases these problems are solved by finding an appropriate 
finite measure by which the energy summations should be implemented. 
For instance, in the Planck hypothesis, which 
concerns the amount of
energy $U(\nu )d\nu$ radiated by the blackbody in the frequency 
range between $\nu$ and $\nu +d\nu$, beside the quantized energy
there is assumed also the existence of a natural
probability which determines the likelihood that the blackbody 
emits-absorbs a certain energy.  
By supposing equal likelihood for radiating $U(\nu )$ (equipartition
principle), the old theory yielded the Rayleigh-Jeans law: 
$U(\nu )d\nu\sim\kappa T\nu^2d\nu$, which
contradicts the empirical curve described by the Wien law: 
$U(\nu )d\nu\sim\nu^3e^{-h\nu/\kappa T}d\nu$. 
The primary evidence justifying the Planck hypothesis was that it 
yielded the Wien law. 

An analysis of Bethe's classical paper, 
written for explaining Lamb shift, 
shows that also his starting formula determining the energy due to 
the interaction of the electron 
with the radiation field assumes equipartition principle.
The applications of cutoff constants and other renormalization 
techniques can be interpreted as a process which turns this summation
into a probabilistic one. It is  
apparent that the solution of this problem requires finding 
an adequate finite complex measure which describes the considered
interaction on a probabilistic background.
This amplitude is explicitly constructed by means of the spectrum of 
$V^{(a,a)}$, which, in its final form, appears in terms of the Gamma 
function $\Gamma (z)$. The natural computation yielding    
the Lamb shift formula in the last section 
is the major evidence which justifies that this amplitude most 
adequately describes the interaction of the electron with the
Coulomb field on the quantum level.

Since this non-perturbative computation leads to the very same 
formula, the renormalization technique seems to have expended 
most of its effort to take off the transmission operators
$V^{(a,b)}$ defined for $a\not =b$ from the Coulomb field and establish,
by means of the remaining $V^{(a,a)}$, an energy summation which has 
probabilistic features. The question arises if
these omitted transmission operators are existing real physical 
objects? Before answering this question note that these 
operators are
of zero trace class satisfying $V^{(a,b)}=\overline V^{(b,a)}$. These 
properties mean that the same amount of energy is transported from
$\mathcal H^{(a)}$ to $\mathcal H^{(b)}$ as from
$\mathcal H^{(b)}$ to $\mathcal H^{(a)}$.
Thus there is no real energy transmission provided by these
operators. What are then these transmission operators for? We suggest
the following answer to this problem: The transmission operators are
real existing parts of the Coulomb operator whose main role is to 
maintain the zonal structure. Without their action all the zones and 
zonal electrons would blow up. This explanation gives a 
satisfactory answer for the most difficult question arising in all
theories which work with extended particles: Why do the
spread-out zonal charged particles not blow up?

\section{Zeeman manifolds}

\subsection{Zeeman-Hamilton operators.} 
The classical Zeeman operator of a charged particle is 
\begin{equation}
\label{Zee_int}
H_Z=-{\hbar^2\over 2m_r}\Delta +
{\hbar eB\over 2m_r c\mathbf i}
D_z\bullet
+{e^2B^2\over 8m_r c^2}(x^2+y^2) +V ,
\end{equation}
where $V=-Z_pe^2/r$ is the Coulomb potential 
originated from the nucleus 
formed by $Z_p$ protons, 
furthermore,
$m_r =m_em_p/(m_e+m_p)$ 
is the reduced mass which differs from the mass, $m_e$, of the 
electron just by a small amount (in case of the hydrogen atom, 
$(m_e-m_r)/m_e\approx 5\times 10^{-4}$). 
Because of the small difference, we use 
$m_e=9.10938188 × 10^{-28} g$ also in the place of $m_r$.

This operator is written up
in Gaussian (centimeter-gram-second) units, where the Bohr magneton
is defined by 
$\mu_B=e\hbar /2m_ec$ and the unit charge, esu of charge or statcoulomb,
is defined so that the Coulomb force constant is 1. 
In SI (m-kg-s) units the Bohr magneton is 
$\mu_B=e\hbar /2m_e = 9.274 009 49(80) × 10^{-24} J·T^{-1}$, where
the magnetic field has SI units of tesla, $1 T = 1 kg·s^{-1}·C^{-1}$
and symbol $C$, Coulomb, denotes unit of electric charge. The
corresponding operator in SI units omits the light speed 
$c=299 792 458\\ m/s$ from the second and third term of (\ref{Zee_int})
and the Coulomb force constant is 
$k_C\approx
8.988×10^9 N\cdot m^2\cdot C^{-2}
$. 
Later computations mostly prefer the SI units. The classical papers 
quoted later use the Gaussian units, however, and this is why the 
units will be used in a mixed way.
  
This operator is usually considered
on the 3-space. The free particle operator restricted onto the 
$(x,y)$-plane (i. e., $V=0$ and $\Delta$ is the Euclidean Laplacian
on $\mathbf R^2$) is called Landau Hamiltonian. This paper proceeds
with this 2D-version and its generalizations   
defined on complex vector spaces
$\mathbf C^{k/2}=\mathbf R^k$.
Operator  
$D_z\bullet =x\partial_y-y\partial_x$, 
called angular momentum operator, 
commutes with the remaining part, 
$
\mathbf O=H_Z-{\hbar eB\over 2m_r c\mathbf i}D_z\bullet
$, 
of the complete Hamiltonian,
thus the spectrum appears on common eigenfunctions. I. e., the
$D_z\bullet$ splits the spectral lines of $\mathbf O$, which phenomena
is associated with the Zeeman effect. 
Actually, the $H_Z$ is the Hamilton operator of an 
electron orbiting about the origin of the 
$(x,y)$-plane in a constant magnetic field $\mathbf K=B\partial_z$.
Since it is still revolving around the origin,
the free Landau particle is only latent-free. 
The 3D-version can be established by means of the Maxwell equations
and the real Heisenberg group representation. To establish the Landau
Hamiltonian, one can use the Fock-Bargmann representation of the
complex Heisenberg group. 

\subsection{Mathematical modeling: Zeeman manifolds.}
Interestingly enough, the Landau operator, $H_Z$, 
can be identified 
with the Laplace operators of two step nilpotent Lie groups endowed 
with the natural left invariant metrics. As far as the author knows,
this interpretation is unknown in the literature. 
A 2-step nilpotent metric Lie group is defined on the
product 
$\mathbf v\oplus\mathbf z$ 
of Euclidean spaces, where the components, 
$\mathbf v=\mathbf R^k$ and
$\mathbf z=\mathbf R^l$, 
are called X- and Z-space respectively. The Lie algebra is completely
determined by the linear space, 
$J_{\mathbf z}$,
of skew endomorphisms acting on the X-space defined by 
$\langle [X,Y],Z\rangle =\langle J_Z(X),Y\rangle$, where 
$X,Y\in\mathbf v$
and $J_Z$ is the endomorphism associated with  
$Z\in\mathbf z$. To be more precise, the Lie algebra is uniquely
determined by the system  
$
\{\mathbf v\oplus\mathbf z, 
\mathcal A : \mathbf z\to SE(\mathbf v)\}
$, 
where 
$ 
\mathcal A : Z\to J_Z
$
is a one-to-one linear map from $\mathbf z$ into the space of skew
endomorphisms acting on $\mathbf v$. If the range,   
$J_{\mathbf z}$, is the same for two maps, then they define isomorphic
Lie algebras which are isometricly isomorphic if the combined map
$
\mathcal A_2^{-1} 
\mathcal A_1 : \mathbf z\to \mathbf z
$
is orthogonal. The natural innerproduct on $\mathbf z$ is defined by
$
\langle Z_1,Z_2\rangle =-TrJ_{Z_1}J_{Z_2}
$.
The metric, $g$, is the left invariant extension of the
natural Euclidean metric on the Lie algebra. The exponential map
identifies the Lie algebra with the group. Thus also the group 
is defined on the same space of $(X,Z)$-vectors on which the Lie 
algebra is living.  

Particular 2-step nilpotent Lie groups are the
Heisenberg-type groups, introduced by Kaplan \cite{k}, 
defined by endomorphism spaces
satisfying the Clifford condition $J^2_Z=-|Z|^2id$. These metric groups 
are attached to Clifford modules, thus the classification
of these modules provides classification also for the H-type groups.
In this case the X-space decomposes into the product 
$\mathbf v=(\mathbf R^{r(l)})^{a+b}=\mathbf R^{r(l)a}\times
\mathbf R^{r(l)b}$ and
endomorphisms $J_Z$ 
are defined by endomorphisms $j_Z$
acting on the smaller space $\mathbf R^{r(l)}$. Namely, the $J_Z$
acts on  
$\mathbf R^{r(l)a}$ resp.
$\mathbf R^{r(l)b}$ as $j_Z\times\dots\times j_Z$ 
resp. $-j_Z\times\dots\times -j_Z$. 
The H-type groups are denoted by 
$H^{(a,b)}_l$, indicating the above decomposition. 

The Laplacians on H-type groups are of the form
\begin{equation}
\label{Delta}
\Delta=\Delta_X+(1+\frac 1{4}|X|^2)\Delta_Z
+\sum_{\alpha =1}^r\partial_\alpha D_\alpha \bullet,
\end{equation}
where $D_\alpha\bullet$ denotes directional derivatives along
the fields 
$J_\alpha (X)=J_{Z_\alpha}(X)$  
and $\{ Z_\alpha\}$ is an orthonormal basis on the Z-space. 
This operator is not the Landau operator yet. It appears, however, on 
center periodic H-type groups,  
$\Gamma\backslash H$, 
defined by factorizing the center of the group
with a Z-lattice $\Gamma =\{Z_\gamma\}$. 
In fact, in this case the $L^2$ function
space is the direct sum of function
spaces $W_\gamma$ spanned by functions of the form
$\Psi_\gamma (X,Z)
=\psi (X)e^{2\pi\mathbf i\langle Z_\gamma ,Z\rangle}.
$
Later on, the Fourier-Weierstrass decomposition 
$
L^2_{\mathbb C}(\mathbb R^k\times T^l)= 
\sum_\gamma W_\gamma
$
is called {\it primary splitting}.
Each $W_\gamma$ is invariant under 
the action of $\Delta$. More precisely, 
$\Delta \Psi_{\gamma }(X,Z)=\Box_{\gamma}\psi (X)
e^{2\pi\mathbf i\langle Z_\gamma ,Z\rangle}$ holds
where operator $\Box_{\gamma }$, acting on 
$L^2(\mathbf v)$, is of the form
\begin{equation}
\label{Box}
\Box_{\gamma }
=\Delta_X + 2\pi\mathbf i D_{\gamma }\bullet -4\pi^2
|Z_\gamma |^2(1 + 
\frac 1 {4} |X|^2).
\end{equation}
When the invariant subspaces are defined by the functions
$\Psi_\gamma (X,Z)
=\psi (X)e^{-2\pi\mathbf i\langle Z_\gamma ,Z\rangle},
$
this operator appears in the following form:
\begin{equation}
\label{Box'}
\Box_{\gamma }
=\Delta_X - 2\pi\mathbf i D_{\gamma }\bullet -4\pi^2
|Z_\gamma |^2(1 + 
\frac 1 {4} |X|^2).
\end{equation}
  
The first basic observation in this paper is that the Landau 
Hamiltonian satisfying $B=1T$ and  
$|e|=1.60217653\times 10^{-19}C$
(elementary charge) can be identified with 
$H_Z=-(\hbar^2/2m_r )\Box_\gamma$, where cases (\ref{Box}) and 
(\ref{Box'})
correspond to $e\,>\,0$ and $e\,<\,0$ respectively. Furthermore, 
$\pi |Z_\gamma |= \lambda=|e|/2\hbar c$ holds, thus the  
$\lambda^2$ can be interpreted as the eigenvalue of $-\pi^2J^2_\gamma$.
Also note that switching the angular momentum endomorphism $J_\gamma$
to $-J_\gamma$ transforms (\ref{Box'}) to (\ref{Box}). This means
that choosing the sign of the charge is equivalent to choosing
$J$ or $-J$ from the set of complex structures available on 
$\mathbf R^2$. Then the operators, 
regarding both for electron and positron,
appear in the common form (\ref{Box}),
where $e$ is a positive quantity.
Later on, this operator is denoted also by $\Box_\lambda$.

Comparing with the Landau operator, this one contains 
a surplus constant $4\pi^2|Z_\gamma |^2=4\lambda^2$, which contribution
does not show up in the original
Landau Hamiltonian. In a physical situation, one considers the 
$\hbar^2/2m_r$-times of this term. Then,
by $m_r\approx m_e,\, \mu_B=2\lambda\hbar^2/2m_e$, 
one has: 
$W_{extra}=(2m_e/\hbar^2)\mu_B^2\approx 1.408970181\times 
10^{-8}kg\cdot s^{-2}T^{-2}$. 
Dimension $kg\cdot s^{-2}=J/m^2$ shows that this extra term is 
a constant energy density field whose integral on the whole space
would provide infinite energy. Unit $T^{-2}$ indicates that this
energy is nothing but the self energy $q\int\mathbf B^2$ of the constant
magnetic field. 
The exact numerical values for the other terms are:
$2\lambda =2m_e\mu_B/\hbar^2\approx 1.492298399
\times 10^{15}m^{-2}T^{-1}$ and 
$\hbar^2/2m_e=6.1042635\times 10^{-39}kg\cdot m^2$. These values are used
in the later computations. 

\subsection{Interpretations for the basic objects}
\label{interpret} 
For a $(k+1)$-dimensional Heisenberg group, defined by a complex
structure $J$ acting on the even dimensional Euclidean space  
$\mathbf v=\mathbf R^k$, 
number $k/2=\kappa$ is interpreted as 
the number of particles. In view of the 2D 
Landau operator, this is the most natural 
interpretation for this number.
Also the interpretation for the X-space
$\mathbf v=\mathbf R^k$ 
is clear. It must be the space where the particles are orbiting in their
own constant magnetic fields such that each particle occupies a complex
plane. These interpretations can be carried over to general center 
periodic 2-step nilpotent Lie groups, where, 
due to the higher dimensional
Z-space, the Zeeman operator appears 
in a more complicated form.

On Heisenberg groups,the center is interpreted as the non-relativistic 
time-axis but this concept can not be obviously taken over to the general
cases. To the complete interpretation of the Z-space one must
answer, first, that how is the time measured
in these models? Equally important question is if 
this model is relativistic 
or non-relativistic?

The key to answering these questions is the realization that
the Lie algebra is uniquely determined by the map
$ 
\mathcal A : Z\to J_Z
$
which corresponds skew endomorphisms to 
the elements of the center. Instead
of time coordinates, this map a priori corresponds angular 
momentums which define constant magnetic fields and vanishing 
electric fields for the particles. But this particular appearance of
a constant electromagnetic field pins down, 
on each complex plane occupied
by a particle, a unique inertia system with
well defined self-time, $t(J_Z)$, measured 
in the system. This denotation 
indicates that this self-time depends on the 
angular momentum endomorphism. 
Also note that the inertia system defined for proportional
$Z$'s are uniquely determined and   
the line spanned
by $Z$ and parameterized by the arc-length $t$ can be interpreted
as the self-time-axis in the inertia system.  

This construction attributing self-times to the particles
relates our model to Dirac's famous multi-time model, where, 
in order to establish a 
relativistic quantum theory, self-time is attributed to the particles.
By this reason, our model can be called relativistic as far as Dirac's
multi-time theory fits this characterization. 
It should be point out that 
this is not the classical relativism and 
for distinguishing from the original theory, this is called 
{\it anchored relativism}. This name was chosen to recall the key idea
in the self-time construction: the 
constant 
electromagnetic field 
$(\mathbf B_\alpha=\mathbf constant\not =0,\mathbf E_\alpha=0)$
defined by an angular momentum endomorphism $J_Z$ ``anchors" 
the system in a unique inertia system which defines the self-time
for a particle.

This self-time construction can be implemented on the center 
of the non-periodic group, after which, 
operator (\ref{Delta}) can be interpreted 
as a positive definite version of the 
Klein-Gordon operator corresponding 
to the anchored relativistic theory. This argument shows that
this anchored relativism does not contradicts the original
relativistic theory, however, it can not be identified with it either.
For instance, 
unless introducing negative energies into the system, 
one can not transform operator (\ref{Delta}) into the original
(indefinite) Klein-Gordon operator just by changing the sign before
$\Delta_Z$ to the minus one. 
An other distinguishing feature is that 
this model leads to probabilistic 
quantum theory working with positive probabilities defined just on the 
space. Recall that Dirac's relativistic electron 
theory (which is different from 
his above mentioned multi-time theory) establishes such positive 
probabilities on the Minkowski space-time. This idea was strongly 
criticized by Pauli, according to whom such probabilistic theory
makes sense only on the space. 

The periodic model can be regarded as a partial crystal model
where the crystal is in the center of the group. The system
can be in crystal states represented by the endomorphisms
$J_{\gamma}$. Parameters $\lambda_i>0$ are defined by the
absolute values of the eigenvalues of
$\pi J_\gamma$ appearing on the corresponding complex eigenplanes. Thus 
$|\pi Z_\gamma |^2=\lambda_1^2+\dots +\lambda^2_\kappa$. 
The Hamilton operators belonging to these crystal states are 
$-{1\over 2}\Box_\gamma$. In case of a single eigenvalue $\lambda$ with
multiplicity $\kappa$, 
the corresponding operator is denoted also by $\Box_\lambda$. Then    
$|\pi Z_\gamma |^2=\kappa\lambda^2$ holds. This paper deals
only with such systems.

\subsection{Isospectrality constructions}

The Riemannian manifolds considered here were
originally used for constructing isospectral
manifolds with different local geometries \cite{sz1}-\cite{sz4}. The
isospectrality examples arouse on certain compact 
submanifolds both of the center-periodic and the 
non-periodic groups. Here only the examples constructed on Heisenberg-type
groups $H^{(a,b)}_l$ will be explained. For a fixed $l$,
what is the dimension of the center, consider all those groups for which
also $(a+b)$ is the same value. All these metric groups live on the same
manifold   
$\mathbf R^{r(l)a}\times\mathbf R^{r(l)b}\times \mathbf R^l$. 
The only difference between
two groups in a family is exhibited by the endomorphisms $J_Z$ which 
are defined by the endomorphisms $j_Z$
acting on the smaller space $\mathbf R^{r(l)}$ such that the $J_Z$
acts on  
$\mathbf R^{r(l)a}$ resp.
$\mathbf R^{r(l)b}$ as $j_Z\times\dots\times j_Z$ 
resp. $-j_Z\times\dots\times -j_Z$. By the physical interpretation,
the $j_Z$ resp. $-j_Z$ correspond to positrons resp. electrons, therefore,
these groups are attached to the same number of particles and the only
distinguishing feature is the ratio of numbers of electrons and positrons.
One can go from one system to the other by exchanging some of the
electrons for positrons.

It is well known in physics that the spectrum does not change during
electron-positron exchanges. A strict convert of this physical statement
to a mathematical one is that the spectra  computed on the whole   
non-compact center-periodic groups $\Gamma\backslash H^{(a,b)}_l$ 
for the members of the considered family are same. In the above papers
this isospectrality is established in a much stronger form, namely, 
not just on the non-compact groups (which cover the physical cases) 
but also on a wide range of compact submanifolds such as
ball$\times$torus-,
sphere$\times$torus-,
sphere$\times$ball-,
sphere$\times$sphere-,
ball-, and sphere-type submanifolds.
Although the physical statement gives some chance for these 
much stronger isospectralities, these statements are rather non-trivial
because a non-trivial isospectrality of two manifolds does not imply
the isospectralities of the submanifolds.

There is an other very surprising statement established in the
above papers. Namely, the members in an isospectrality family can have
different local geometries. This statement is true, for 
instance,
exactly for those families of Heisenberg-type groups where number $l$ 
is of the form $l=4r+3$, where $r=0,1,\dots$. The other families
consist of isometricly isomorphic groups for which the isospectrality
is a trivial statement. 

We demonstrate the different local geometries for the members
of the isospectrality family
$H^{(a,b)}_3$,
where the Z-space, $\mathbf R^3$, is identified with the space of
imaginary quaternionic numbers, the ``small'' X-space is
$\mathbf R^{r(l)}=\mathbf R^4=\mathbf H$ (the space of quaternions), 
and the endomorphisms are defined by $j_Z(h)=Zh$. 
When the group is attached to the same type of
particles (i.e., it is
$H^{(a+b,0)}_3$ or
$H^{(0,a+b)}_3$),
then both the
sphere$\times$torus- and
sphere$\times$sphere-type
submanifolds are homogeneous (the isometries act transitively), while
for mixed particles, characterized by the relation $ab\not =0$, these
submanifolds are locally inhomogeneous. The existence of
isospectral metrics having different local geometries is 
unknown even in physics where no geometries attached to spectra have
been considered so far. This phenomenon 
certainly must have some effect on a  
deeper understanding of the symmetries on the quantum level, but this
impact is not well understood yet.

\section{Normal de Broglie Geometry}
\subsection{ Introducing the zones.} 
In what follows, all investigations are performed on the X-space.
Actually, the following parts are understandable without knowing
about the mathematical model described above and it can be considered
as spectral analysis of the Zeeman operator. It is connected with
the above model such that the invariant subspaces $W_\gamma$ defined
by the primary decomposition are further investigated, which, by the
map $\Psi_\gamma (X,Z)\to \psi(X)$, can be identified with the space 
$L^2_{\mathbf C}(X)$ 
consisting functions defined just on the X-space. 
The Laplacian $\Delta$ appears on a fixed invariant subspace as the
Landau-Zeeman operator $\Box_\lambda$ which acts on complex valued 
functions defined on the X-space. The zones are defined by a spectral
decomposition of 
$L^2_{\mathbf C}(X)$,
thus it can be regarded as a secondary decomposition of the subspaces
obtained in the primary decomposition. 
In this paper the endomorphism  
$-\pi^2J^2_\gamma$ 
has only one eigenvalue $\lambda^2$ with multiplicity $k$, thus 
$|\pi J_\gamma |^2=\kappa\lambda^2$ holds.  

The Hilbert space, $\mathcal H$, of the complex valued
$L^2$-functions is isomorphic to the weighted space defined by 
the Gauss density $d\eta_\lambda (X)=e^{-\lambda |X|^2}dX$. The latter 
space is spanned by the complex valued polynomials. 
Next $\mathcal H$ is considered in this form.
The natural {\it complex Heisenberg group 
representation} on $\mathcal H$ is defined by  
\begin{equation}
\label{rho}
\rho_{\mathbf c} (z_i)(\psi )=(-\partial_{\overline{z}_i}+
\lambda z_i\cdot )\psi\quad ,\quad
\rho_{\mathbf c} (\overline{z}_i)(\psi )=\partial_{z_i}\psi , 
\end{equation}
where $\{z_i\}$ is a complex coordinate system on the X-space.
This representation is reducible. In fact, it is irreducible on the
Fock space generated by the holomorphic polynomials, 
where it is called Fock-Bargmann representation.
Besides the Fock space there are infinitely many
other irreducible invariant subspaces. By this reason, the above 
representation is called {\it extended Fock-Bargmann representation}
and the irreducible decomposition defined by this representation is called
{\it secondary splitting}. 
In the function operator
correspondence, this representation associates operator (\ref{Zee_int})
to the Hamilton function of an electron orbiting in a constant 
magnetic field.

The zones are defined in two different ways. First, they can be defined
by the invariant subspaces of representation (\ref{rho}). The actual 
construction uses Gram-Schmidt orthogonalization. On the complex plane
$\mathbf v=\mathbf C$, corresponding to the 2D Landau operator, 
the $\mathcal H$ is the direct sum of subspaces 
$G^{(a)}$ spanned by functions of the form $\overline z^ah$, where
$h$ is an arbitrary holomorphic polynomial. Then one gets the zones
$\mathcal H^{(a)}$, where $a=0,1,2,\dots$, 
by the Gram-Schmidt orthogonalization process applied to the 
function spaces $G^{(a)}$. 
It is clear that the first zone, $\mathcal H^{(0)}$, is the Fock 
space.  The zone index $a$ indicates the maximal number
of the antiholomorphic coordinates $\overline z$ in the polynomials
spanning the zone. 

One of the referees of \cite{sz6} pointed out to me that the 
polynomials produced by this constructions were considered also 
by It\^o \cite{i} in the context of complex Markov processes.
In fact, It\^o defines the Hermite polynomials of complex variables
for $p,q=0, 1, 2,\dots$ by the explicit formula
\begin{equation}
H_{pq}(z,\overline z)=\sum_{s=0}^{min(p,q)}(-1)^s{p!
q!\over s! (p-s)!(q-s)!}
z^{p-s}\overline z^{q-s}. 
\label{ito}
\end{equation}
In the 2D case, they form an orthogonal basis in 
$\mathcal H$ defined for $\lambda =1$. In this formalism, the zones are
spanned by polynomials belonging to fixed values of $q$, i. e., the
$q$ corresponds to the zone index $a$ in our formalism.
Corresponding to the cases $p\geq q$ resp. $q\geq p$, 
these formulas appear in the form 
$f_n(r^2)z^{p-q}$ resp. 
$f_n(r^2)\overline z^{q-p}$,
where $f_n(t)$ is a polynomial of order $n=min(p,q)$. 
These formulas 
will be reconstructed in Section \ref{coulomb}
in terms of the Laguerre polynomials
\begin{equation}
\label{rec0}
L^{(l)}_n(t)=\sum_{i=0}^n{n+l\choose n-i}{(-t)^i\over i!},
\end{equation}
where they appear in the form 
\begin{equation}
\label{regeigf}
(-1)^nn!L^{(l)}_n(r^2)z^l\quad \mathrm{resp.}\quad
(-1)^nn!L^{(l)}_n(r^2)\overline z^l.
\end{equation}
Then, substitution $r^2=z\overline z$ converts (\ref{regeigf}) to 
(\ref{ito}).

The construction with the Gram Schmidt orthogonalization 
easily extends to general dimensions. Gross zone $\mathcal H^{(a)}$ is
constructed by means of all polynomials 
$
\overline z_{1}^{(a_1)} \dots
\overline z_{k/2}^{(a_{k/2})} 
$
satisfying $a_1+\dots +a_{k/2}=a$. This gross zone is the direct
sum of the subzones $\mathcal H^{(a_1\dots a_{k/2})}$ defined 
for the particular values 
$a_1,\dots ,a_{k/2}$. The eigenfunctions appear as an appropriate
product of eigenfunctions defined for the complex coordinate planes
(these details are mostly omitted in this review, but see further
remarks in the end of this section). 
In the 2D-case all the zones are irreducible under the 
action of the extended Fock-Bargmann representation. In the higher 
dimensions, however, the {\it gross zones} are reducible and the 
subzones are irreducible. Note that the holomorphic (Fock) zone 
is always irreducible. For the sake of simplicity, 
all the formulas below are established on the gross zones.

In terms of Ito's polynomials (\ref{ito}), 
which are defined for arbitrary
dimension $k$, the
eigenfunctions appear in the form
$
h^{(p,\upsilon)}(X)=
H^{(p,\upsilon)}(X) 
e^{-\lambda |X|^2/2}
$
with the corresponding eigenvalues
$
-((4p+k)\lambda +2k\lambda^2),  
$
where $p$ resp. $\upsilon$ are the holomorphic resp. antiholomorphic
degrees of polynomial $H^{(p,\upsilon )}$ (the last term is due to
to the energy-density defined by the constant magnetic field where 
$4(k/2)=2k$ applies). 
Numbers $\tau =p+\upsilon$ and $m=2p-\tau$ are
called {\it total-} and {\it magnetic quantum numbers} 
(TQN and MQN) respectively. The
above function is an eigenfunction also of the magnetic dipole
moment operator with eigenvalue $m$.
Then a zone is spanned by 
eigenfunctions having the same index $\upsilon$. 
According to the formula $\upsilon ={1\over 2}(\tau -m)$, 
the zones are determined by the quantum numbers $\tau$ and $m$.
For a given $\tau$, the range of $m$ is $-l,-l+1,\dots,l-1,l$,
where $l=|p-\upsilon|$,
and the eigenfunctions belonging to different MQN's are sorted out to
distinct zones.
In this sense, a zone exhibits the magnetic state of the particle.
Note that eigenvalues are independent of the 
antiholomorphic index and they depend just on the holomorphic index. 
As a result, each eigenvalue has infinite multiplicity. 
On the irreducible zones, however,
each multiplicity is $k/2=\kappa$. Moreover, two irreducible zones 
are isospectral.

It is important to understand that the above spectrum computation is 
not the standard one, in which case the eigenfunctions 
are sought in the form
$f_{n,l}(|X|^2)G^{(l)}(X)e^{-{1\over 2}\lambda |X|^2}$, where
$G^{(l)}$ is an $l^{th}$-order homogeneous harmonic polynomial and 
$f_{n,l}(t)$ is an $n^{th}$-order polynomial 
which depends also on $l$. These standard explicit eigenfunction 
computations are completely established in Section\ref{expeigf}.
It was Schr\"odinger who computed the eigenfunctions and eigenvalues
of his operators in this form. The classical quantum numbers are 
derived from this representation of the eigenfunctions.
Note that formulas (\ref{regeigf}) represent the eigenfunctions exactly
in this form.
Indeed, in the 2D case, the homogeneous harmonic
polynomials are of the form $z^l$ or $\overline z^l$ 
and functions $f_{n,l}$ happen to be the corresponding Laguerre
polynomials. 
This conversion is such easy just in the 2D-case. In \cite{sz6}, 
these computations are established for arbitrary dimensions, where
2 different type of eigenfunction computations are developed. One of them
is of traditional (Schr\"odinger) type and the other seeks the 
eigenfunctions as the product of It\^o's polynomials.   
According to the traditional terminology of spectroscopy, 
the azimuthal quantum number is defined by the order, $l$, of $G$
and order $n(l)$ is the radial quantum number. 
Thus $\tau =l+2n(l)$ and $p=n+l, \upsilon=n$ resp. $p=n,\upsilon =n+l$
hold, corresponding to the cases indicated in (\ref{regeigf}).
The magnetic
quantum numbers, $m$, are defined in both cases by the same numbers.

These formulas clearly describe the conversion of quantum numbers
defined by the different representation of the eigenfunctions. However,
the standard representation with radial functions and spherical 
harmonics eclipses the zonal structure even in the 2D case, which is, 
on the other hand, very clearly exhibited by the new type of 
technique also in general dimensions. In the zonal theory the It\^o
polynomial technique and not the standard one is the natural tool
for developing a clear spectral analysis. This preference refers also
to the quantum numbers defined by the two techniques.   
Probably these
standard computational techniques constitute the main reasons for the
zones have not been investigated in the literature earlier. Nor have
the intriguing fact, asserting that functions (\ref{ito}) are 
eigenfunctions of the Landau Hamiltonian, been exploited so far.

\subsection{Projection kernels and point-spreads.} In the literature 
only the projection onto the Fock space 
$\mathcal H^{(0)}$ is well known, which turned out to be
a convolution operator with the so called Fock-Bargmann kernel
$
\big({\lambda \over\pi}\big)^{k/2}
e^{\lambda (z\cdot\overline{w}-{1\over 2}(|z|^2+ |w|^2)}.
$ 
Our theory, developed in \cite{sz5,sz6}, explicitly determines 
the projection also onto a general zone $\mathcal H^{(a)}$. 
Then the corresponding projection kernel is 
\begin{equation}
\delta_{\lambda z}^{(a)}(w)
= \big({\lambda \over \pi} \big)^{k\over2}
L^{({k\over 2}-1)}_a(\lambda |z-w|^2)
e^{\lambda (z\cdot\overline{w}-{1\over 2}(|z|^2+ |w|^2)},
\end{equation}
where $L_a^{((k/2)-1)}(t)$ is the Laguerre polynomial indicated by the
indexes. To have this formula, consider an orthonormal basis
$\{\varphi_i^{(a)}\}_{i=1}^\infty$ formed by eigenfunctions being in 
$\mathcal H^{(a)}$. The projection kernel can be
formally expressed in the form
$
2\delta^{(a)}(z,w)=\sum_i \varphi^{(a)}_i(z)
\overline{\varphi^{(a)}_i(w)},
$
where $z$ and $w$ represent complex vectors on 
$\mathbf C^{\kappa}=\mathbf R^k$. Then the formula
can be established by means of the explicit eigenfunctions.
These kernels can be interpreted as restrictions of the 
global Dirac delta distribution, 
$2\delta_z(w)=\sum \varphi_i(z)\overline{\varphi}_i(w)$, onto the zones.

These kernels represent one of the most important concepts
in this theory. They can be interpreted such that, 
on a zone, a point particle appears as a spread
described by the above wave-kernel. 
Note that how these kernels, called zonal point-spreads, are derived
from the one defined for the holomorphic (Fock) zone. This holomorphic 
spread, which involves a Gauss function, is just multiplied by the 
radial Laguerre polynomial corresponding to the zone. This form of
the functions describing the point-spreads
show the most definite similarity to the de Broglie wave packets. 
In a rigorous theory, function 
$
\delta^{(a)}_{\lambda Z}
\overline{\delta}^{(a)}_{\lambda Z}
$
is the density of the point-spread concentrated around $Z$ and 
$
\delta^{(a)}_{\lambda Z}
$
is the so called spread-amplitude.
On a given zone the point-spreads are the
most compressed wave packets, yet they are distributed all over the
whole space. This zonal particle theory gives a clear explanation
for the Aharanov-Bohm (AB) effect \cite{ab} as well as other 
phenomenas described in \cite{sz5}.

{\it The AB effect} produces
relative phase shift between two electron beams enclosing a magnetic
flux even if they do not touch the magnetic field. This
effect has no explanation in the classical mechanics and it
contradicts even the relativistic principle of  
{\it all fields must interact only locally}. Yet, this effect was 
clearly demonstrated by the {\it Tonomura et al experiments} 
\cite{t1,t2}.

Although the point electrons do not touch the fields, the vector 
potential involved into the Hamilton operator of the system does reach 
there. Exploiting this phenomena, Aharanov and Bohm explained the effect
by the ``significance of electromagnetic potentials in the quantum
theory". In classical physics this potential is considered to be 
a mere mathematical convenience which is completely meaningless from 
physical point of view. In de Broglie geometry the
zonal particles are extended ones which must touch the magnetic
field, which is a clear enough 
explanation for the AB effect. Since the zones are defined by a 
particular vector potential, this explanation is in accordance with 
the Aharanov-Bohm idea.
Indeed, the vector potential is not just a mathematical 
convenience any more but it is one of the important
physical objects by which the zonal structure is defined. 

Despite that the experiments were performed 
under the condition of complete confinement of the magnetic field 
in the magnet, some physicists have questioned the validity of the 
tests, attributing the phase shift to leakage fields. The electron 
spread idea developed in this paper can be interpreted such that 
not the magnetic field but ``the electrons are leaking".

\subsection{Global Wiener- and Schr\"odinger-flows.}
The zonal analysis is continued in this section by describing the
global flows defined on the total Hilbert space $\mathcal H$.
Because of its trivial contribution to the formulas, the surplus
constant belonging to the constant magnetic field is omitted and 
$H_Z$ means the
Landau Hamiltonian. The global Wiener-flow,
$
e^{-tH_Z}(t,X,Y),
$ 
appears in the following explicit form:
\begin{eqnarray}
\label{d_1gamm}
\big({\lambda\over 2\pi sinh(\lambda t)}\big)^{k/2}e^{
-{\lambda}({1\over 2}coth(\lambda t)
|X-Y|^2+\mathbf i
\langle X,J (Y)\rangle}.
\end{eqnarray}
This kernel satisfies the Chapman-Kolmogorov identity and it tends
to $\delta(X,Y)$ when $t\to 0_+$. However, 
it is not of the trace class, thus functions such as 
the partition function or the zeta function are not defined in the
standard way. Also note that by regularization (renormalization) only
well defined relative(!) partition and zeta functions are introduced.

The global Schr\"odinger kernel,
$
e^{-t\mathbf iH_Z}(t,X,Y),
$ 
appears in the following explicit form:
\begin{eqnarray}
\big({\lambda\over 2\pi
\mathbf i sin(\lambda t)}\big)^{k/2}e^
{\mathbf i{\lambda}\{
{1\over 2} cot(\lambda t)|X-Y|^2-\langle X,J(Y)\rangle \}}.
\end{eqnarray}
Since for fixed $t$ and $X$ the function depending on $Y$ is 
not $L^2$, the integral required
for the Chapman-Kolmogo\-rov identity is not defined for this kernel. 
Neither is this kernel of the 
trace class. Nevertheless, it satisfies the 
above limit property when $t\to 0_+$. 

It is well known that rigorously defined
measure on the path-spaces can be introduced only with the Wiener
kernel $e^{-tH}$.  
Note that the heat kernel involves a Gauss density which makes this
constructions possible. Whereas, the Schr\"odinger 
kernel does not involve
such term. This is why no well defined constructions can be
carried out with this kernel. These
difficulties disappear, however, by considering these 
constructions on the zones separately.

\subsection{Zonal Wiener- and Schr\"odinger-flows.} The zones are 
invariant with respect to the action of the Hamilton (Laplace) 
operator, thus the zonal flows are well defined on each zone. 
The zonal
Wiener-kernels are of the trace class, which can be
described by the following explicit formulas. 
\begin{eqnarray}
\label{d_1^a}
e^{-tH_Z^{(0)}}=
\big({\lambda e^{-\lambda t}\over \pi}\big)^{k\over 2}
e^{ \lambda(-{1\over 2} (|X|^2+|Y|^2)+ e^{-2\lambda t}
\langle X,Y+\mathbf iJ(Y)\rangle )},
\\
e^{-tH_Z^{(a)}}=
\mathcal L^{({k\over2}-1)}_a
(t,X,Y))
e^{-tH_Z^{(0)}}(t,X,Y),
\end{eqnarray}
where 
$
\mathcal L^{({k\over2}-1)}_a
$
can be explicitly computed in terms of the corresponding Laguerre 
polynomial and $e^{-2t}$. Furthermore, for the zonal partition function,
$
Tre^{-tH_Z^{(a)}},
$ 
we have
\begin{eqnarray}
\mathcal Z_1^{(a)}(t)
={a+(k/2)-1\choose a}e^{-{k\lambda t\over 2}}/
(1-e^{-2\lambda t})^{k\over 2}.
\end{eqnarray}

Also the zonal 
Schr\"odinger kernels are of the trace class which, together with their
{partition functions}, can be described 
by the following explicit formulas.
\begin{eqnarray}
\label{d_i^a}
e^{-t\mathbf iH_Z^{(0)}}=
\big({\lambda e^{-\lambda t\mathbf i}\over \pi}
\big)^{k\over 2}
e^{ \lambda_i(-{1\over 2} (|X|^2+|Y|^2)+ 
e^{-2\lambda_i t\mathbf i}
\langle X,Y+\mathbf iJ(Y)\rangle )},
\\
e^{-t\mathbf iH_Z^{(a)}}=
\mathcal L_{\mathbf ia}^{({k\over2}-1)}
(t,X,Y))
e^{-t\mathbf iH_Z^{(0)}}(t,X,Y),\\
\mathcal Z_{\mathbf i}^{(a)}(t)
={a+(k/2)-1\choose a} e^{-{k\lambda t\mathbf i\over 2}}/
(1-e^{-2\lambda t\mathbf i})^{k\over 2}
\end{eqnarray}
The zonal Schr\"odinger-kernels are zonal fundamental solutions 
of the Schr\"o\-dinger equation. They
satisfy the Chapman-Kolmogorov identity and tend to
$\delta^{(a)}$ when $t\to 0_+$.
 
On the zones the Wiener and Schr\"odinger kernels are not just 
of the trace class but both define 
complex zonal measures, namely
the {zonal Wiener measure} 
$dw_{1xy}^{T(a)}(\omega)$ 
and the {zonal Feynman measure}
$dw_{\mathbf ixy}^{T(a)}(\omega)$,
on the space of continuous curves $\omega :[0,T]\to \mathbf R^k$ 
connecting two points $x$ and $y$ rigorously.
The existence of zonal Wiener measures
is not surprising. This measure exists even for the 
global setting.
However, the trace class property is a new feature, indeed. In
case of the zonal Feynman measure both the trace class property 
and the existence of the rigorously defined zonal Feynman measures 
are new features. 
Note that also the zonal Schr\"odinger kernels
involve a Gauss kernel which makes these constructions well defined.

\subsection{The non-periodic zones defined by 
Fourier-averaging.\label{nonperiodic}}
This paragraph sketchily describes the construction of zones in
the non-periodic case. (This case is not considered in the rest part
of the article thus all those details are understandable without
this section.)
On center periodic 2-step nilpotent Lie groups the invariant
subspaces $W_\gamma$, defined for a lattice point $Z_\gamma$ by
functions of the form
$\Psi_\gamma (X,Z)
=\psi (X)e^{2\pi\mathbf i\langle Z_\gamma ,Z\rangle},
$
is identified, by the map $\Psi_\gamma\to\psi$, with function space
$\mathcal H$ consisting of functions depending just on the
X-variable. Although the zonal decomposition is established on 
$\mathcal H$, it depends on $\gamma$ and it lives, actually, on 
$W_\gamma$. By considering this zonal decomposition on each
$W_\gamma$, it lives on $L^2(\Gamma\backslash H)$.

Such simple reduction to the X-space is not possible on non-periodic
groups. Unlike in the periodic case, where the zonal functions
involve just one function, 
$
e^{2\pi\mathbf i\langle Z_\gamma ,Z\rangle},
$
which depend on the Z-variable, the zonal functions in a zone on
the non-periodic manifolds involve all the functions which depend
on the Z-variable. Next we describe this construction just on
the H-type groups.

In the first step, for any unit vector $V_u$ of the Z-space, 
consider a complex orthonormal basis 
$\{Q_{V_u1},\dots ,Q_{V_uk/2}\}$
on the complex X-space defined by the complex structure $J_{V_u}$
which defines the complex coordinate system
$\{z_{V_u1},\dots ,z_{V_uk/2}\}$ on the X-space.
This basis field must be smooth on an everywhere dense open subset of
the unit Z-sphere such that it 
is the complement of a set of 0 measure. For given values
$a_1,\dots ,a_{k/2}$  
satisfying $a_1+\dots +a_{k/2}=a$ consider the zone,
$\mathcal H^{(a_1\dots a_{k/2})}_{V_u}$, 
defined by 
$
\overline z_{V_u1}^{(a_1)} \dots
\overline z_{V_uk/2}^{(a_{k/2})} 
$
by the Gram Schmidt orthogonalization.
Then the straight zone, 
$\mathcal S^{(a_1\dots a_{k/2})}$,
is spanned by functions of the form
$
\int_{\mathbf R^l} e^{\mathbf i\langle Z,V\rangle}
\phi (V)
 h^{(a_1\dots a_{k/2})}_{V_u} 
dV,
$
where $\phi (V)$ is an $L^2$-function defined on the Z-space 
$\mathbf R^l$ and 
$
 h^{(a_1\dots a_{k/2})}_{V_u} 
$ is eigenfunction (It\^o-function in a general sense)
from the corresponding zone 
$\mathcal H^{(a_1\dots a_{k/2})}_{V_u}$.

It can be shown that the $L^2$ Hilbert space on the whole group $H$
is the direct sum of the straight zones 
$\mathcal S^{(a_1\dots a_{k/2})}$. The spectral investigations on 
these zones are much more complicated then on the zones defined for
center periodic groups. For indicating the difficulties we mention
that the eigenfunctions of the Klein-Gordon Laplacian $\Delta$ are
of the form 
$
\oint_{S_{R_Z}} e^{\mathbf i\langle Z,V\rangle}
\phi (V)
 h^{(a_1\dots a_{k/2})}_{V_u} 
dV,
$
where $S_{R_Z}$ is a sphere of radius $R_Z$ around the origin of the
Z-space and  $\phi (V)$ is an $L^2$-function defined on this sphere. 
This formula shows that the spectrum of the operator is continuous
and each eigenvalue has infinite multiplicities.
The spectral analysis with such a complicated spectrum will be 
developed elsewhere.

\subsection{Infinities in Quantum Electrodynamics.} 
The problem of infinities (divergent integrals), which has been with us 
since the early days both of quantum field theory (cf. Heisenberg-Pauli 
(1929-30)) and elementary particle physics
(cf. Oppenheimer (1930), Waller (1930)),
is treated by {\it renormalization} in the current theories. 
This perturbative tool provides
the desired finite quantities by differences of infinities. 
The problem of infinities is the legacy of controversial 
concepts such as {\it point mass}
and {\it point charge} of classical electron
theory, which provided the first warning that a point 
electron will have infinite electromagnetic
self-mass: the mass $e^2/6\pi ac^2$ for a surface distribution
of charge with radius $a$ blows up for $a\to 0$. 
Infinity appears also as infinite electromagnetic
energy $\int(\mathbf E^2+\mathbf B^2)dp/8\pi$ of the Coulomb 
electric field $\mathbf E$.
The infinities, related to the divergence of the summations
over all possible distributions of energy/momentum of the virtual 
particles, mostly appear in the form of infinite
traces of kernels such as the Wiener-kernel $e^{-tH}$
or the Schr\"odinger kernel $e^{-tH\mathbf i}$.
 
The basic idea in the new non-perturbative approach presented in this 
paper is that the total quantum Hilbert 
space is broken up into invariant subspaces (zones) which become, 
so to speak, the ``homes" for the zonal particles living there. The
theory investigating these invariant subspaces is called de Broglie 
geometry.  This terminology is chosen to suggest that a point, 
$x$, is a non-existing object on a zone. Rather it appears as 
a point spread defined by projecting the Dirac delta, $\delta_x$, 
onto the zone. I. e., a point becomes a wave packet on a zone whose 
explicit form exhibits its very close kinship to the de Broglie waves.  
In a sense, de Broglie geometry ostracizes the infinities 
by exchanging the points for wave packets and, therefore, 
compels the particles to be extended.

The zones are established by means of the Landau-Zeeman operator.
The biggest challenge, the Coulomb operator, has not appeared 
on the scene yet. By not leaving them invariant, it
actually destroys the zones. This phenomenon requires a completely
new attitude which is developed in the following sections.  

\section{Interaction with the Coulomb field\label{coulomb}}

Operator defined by multiplication with the Coulomb
potential function does
not commute with the rest part (Landau operator) of the complete
Zeeman operator but induces transmissions between the
zones. The transmission from a zone into itself is called {\it zonal
Coulomb operator}. For the explicit description of these transmission
operators one should describe the eigenfunctions of the
Landau operator as well as the action of the Coulomb 
operator on them.

\subsection{Explicit eigenfunctions.}
\label{expeigf}
One can trace back 
the eigenvalue problem of $\Box_{\lambda}$ to the eigenvalue problem
of an ordinary differential operator acting on the radial functions
$f (\lambda\langle X,X\rangle)$ as follows.  
First let the simplest case satisfying $\lambda =1$ be considered.
By $D_{\lambda }\bullet f=0$, $|Z_\lambda |^2=k/2$, and 
$|J_\lambda (X)|^2=\langle X,X\rangle$ we get
\begin{eqnarray}
(\Box_{\lambda}F)(X)=\big(4\langle X,X\rangle f^{\prime\prime}
(\langle X,X\rangle )
+(2k+4\tilde l)f^\prime (\langle X,X\rangle )\\
-(2m +4((1 +{1\over 4}\langle X,X\rangle )
f(\langle X,X\rangle ))\big)H^{(\tilde l,m)} (X).
\nonumber
\end{eqnarray}
The eigenvalue problem is reduced, therefore, to the ordinary
differential operator
$
(L_{(\lambda =1,\tilde l,m )}f)(t)
$
defined by
\begin{equation}
\label{Lf}
4tf^{\prime\prime}(t)
+(2k+4\tilde l)f^\prime (t)
-(2m +4({k\over 2} +
{1\over 4}t))f(t).
\end{equation}
 
The function 
$e^{-{1\over 2}t}$ 
is an eigenfunction of this
operator with eigenvalue $-(4\tilde p+3k)$. The general
eigenfunctions are sought in the form
$
f(t)=u(t)e^{-{1\over 2}t},
$
which is an eigenfunction of 
$L_{\tilde l,m}$
if and only if $u(t)$ is an eigenfunction of operator
$
P_{(\lambda =1,\tilde l,m )}
$
defined by
\begin{equation}
\label{Pop}
4tu^{\prime\prime}(t)
+(2k+4\tilde l-4t)u^\prime (t)
-(4\tilde p+3k)u(t).
\end{equation}

This operator has a
uniquely determined polynomial eigenfunction
\begin{equation}
\label{lag}
u_{(\lambda =1,n,\tilde l,m )}(t)
=t^n+a_1t^{n-1}+a_2t^{n-2}+\dots +a_{n-1}t+a_n
\end{equation}
with coefficients satisfying the recursion formulas
\begin{equation}
a_0=1\quad ,\quad a_i=-a_{i-1}(n-i)(n+\tilde l+{1\over 2}k+1-i)n^{-1}.
\end{equation}
One can easily establish explicit combinatorial formula for $a_i$ by
this recursion. 
The eigenvalue corresponding to this polynomial is
\begin{equation}
\label{leigv}
\mu_{(\lambda =1,n,\tilde l,\nu )}=-(4n+4\tilde p+3k),\,\,
\textrm{where}\,\, \tilde p={1\over 2}(m +\tilde l).
\end{equation}

Polynomials (\ref{lag}) are nothing but 
the Laguerre polynomials, which can be defined by the $n^{th}$-order 
polynomial eigenfunctions of the differential operator
\begin{equation}
\label{Lop}
\Lambda_{\alpha} (u)(t)=tu^{\prime\prime}+(\alpha +1-t)u^\prime ,
\end{equation}
with eigenvalues $-n$.
Therefore
\begin{equation}
P_{(\lambda =1,\tilde l,m )}=4\Lambda_{({1\over 2}k+
\tilde l-1)}-(4\tilde p+3k).
\end{equation}
Thus the eigenfunctions of operators (\ref{Pop}) and (\ref{Lop}) 
are the same indeed. 
Particularly we get that, for fixed values of $k,\tilde l,m$ 
and $\tilde p$, the functions 
$
u_{(\lambda =1,n,\tilde l,m )}\, ,\, n=0,1,\dots \infty
$, form a basis in $L^2([0,\infty ))$.

For a single $\lambda$, 
the eigenfunctions are of the form
\[
u(\lambda\langle X,X\rangle )e^{-{1\over 2}\lambda 
\langle X,X\rangle}H^{(\tilde l,m)}(\lambda^{1\over 2}X),
\]
where $u$ corresponds to $\lambda =1$.
The corresponding eigenvalue is then
\begin{equation}
\mu_{(\lambda,n,\tilde l,m)}=-((4n+4\tilde p+k)\lambda+2k\lambda^2),
\end{equation}
which statement is due to the fact that substitution 
$Y=\lambda^{1\over 2}X$ transforms operator 
$\Box_{\lambda ,X}$ to
$\lambda\Box_{\lambda=1 ,Y}$. 

General eigenfunctions defined for a
system $\{\lambda_1,\dots ,\lambda_r\}$ of eigenvalues are the
products of eigenfunctions determined for individual $\lambda_i$'s.
The above explicit formulas can be established by means of these
explicit eigenfunctions in the most general cases. In the 2D-case,
these eigenfunctions appear in the form described in (\ref{rec0})
and (\ref{regeigf}), where $l=\tilde l$ and $p=n+\tilde p$. 
The corresponding polynomials were introduced
also by It\^o, in the context of complex Markov processes, in the
form (\ref{ito}). It was indicated earlier that our theory prefers 
the It\^o polynomial technique, where the eigenfunctions are represented
by products of It\^o's polynomials, to the standard one
in developing an effective zonal spectral investigation. For instance,
in this new representation of an eigenfunction, the functions 
belonging to the same coordinate clearly visualize the corresponding
particle in the system, while they are completely hidden in the standard
representation.  
Next we proceed with the investigations in the 2D-case.

\subsection{Transmissions and fluctuations; 
Zonal Coulomb fields.}

By one of the definitions, the 2D zones are introduced  by means of the 
Zeeman-Landau operator (free particle) 
which omits the Coulomb potential $V=Z_pe^2r^{-1}=Qr^{-1}$ 
due to the nucleus. The 2D Landau operator is defined by a simple 
restriction of
the free 3D Zeeman operator onto the $(x,y)$-plane, therefore, for 
bounded particles, also the Coulomb operator is defined by restricting the 
above 3D potential onto the $(x,y)$-plane. 
This is different from the 2D Coulomb potential, $Q\ln r$, which could
also be considered in these investigations.
The more precise arguments supporting the usage of the 3D-potential 
over the 2D-one are explained in the end of this section.  

First note that the zones are not invariant with respect 
to multiplication with $V$. This Coulomb operator 
induces transmission integral operators
$
V^{(a,b)}: \mathcal H^{(a)} \to\mathcal H^{(b)}
$ 
with smooth $L^2$-kernels
\begin{equation}
V^{(a,b)}(v,w)=\int\delta_\lambda^{(b)}(v,z)V(z)
\delta_\lambda^{(a)}(z,w)dz
\end{equation} 
satisfying $V^{(a,b)}=\overline V^{(b,a)}$.
Transmission operator, $V^{(a,a)}$, mapping the zone $\mathcal H^{(a)}$
onto itself is called zonal Coulomb potential. 
Fluctuation operator on $\mathcal H^{(a)}$
through $\mathcal H^{(b)}$ is defined by  
$
F^{(a\to b\to a)}:=
V^{(b,a)}\circ V^{(a,b)}
$. 
A remarkable feature of the zonal Coulomb potential is that it turns 
the global
interaction exhibited in the ``total" Coulomb law into a local one.
The same statement is valid for the zonal transmission and fluctuation
operators.

In order to understand these operators more deeply, first,
the matrix both of the Coulomb- and the complete Zeeman-operator in
the basis formed by the eigenfunctions (\ref{ito})-
(\ref{regeigf}) will be explicitly described. These eigenfunctions can
be parametrized by the pairs $(p,q=\upsilon )$ 
of zonal quantum numbers, or,
by the classical quantum numbers $(n,m=p-q)$. The corresponding total, 
$\tau =p+q$, and azimuthal, $l=|p-q|$, quantum numbers  are determined
by these ones. Since $V$ is a radial function, integral 
$
\int H_{pq}(z)V(\sqrt{z\overline z})\overline H_{p^\prime q^\prime}(z)dz
$
can be non-zero only for functions defined by the same magnetic quantum
number $m=p-q=p^\prime -q^\prime$. Also note that for functions 
$H_{nm}$, when they are
defined by $m\geq 0$, the $n$ is equal to the zone index
$q=\upsilon$, while, $q=\upsilon =-m+n$ holds in case of $m<0$.
  
Eigenfunctions $H_{0m},H_{1m},H_{2m},\dots$ considered for a fixed
magnetic number $m$ span the so called {\it magnetic subspace}
$\mathcal M_m$. Then, corresponding to $m\geq 0$ resp. $m<0$, 
$\mathcal H^{(a)}\cap\mathcal M_m=\mathbf H_{am}$ resp. 
$\mathcal H^{(a)}\cap\mathcal M_m=\mathbf H_{(a+m)m}$
hold, where $\mathbf H_{nm}$ is the subspace spanned by $H_{nm}$.
Note that $\mathbf H_{(a+m)m}=\mathbf 0$ holds for $(a+m)<0$. 
By the above argument,
the magnetic subspaces are invariant under the action of the
Coulomb operator meaning that the magnetic quantum number is invariant
under the transmissions defined above. Particularly, eigenfunctions
(\ref{ito})-(\ref{regeigf}) are eigenfunctions both of the zonal
Coulomb operators $V^{(a,a)}$ and the fluctuation operators
$V^{(b,a)}\circ V^{(a,b)}=F^{(a\to b\to a)}$. 
Thus, both commute with
the Landau operator $\Box_{\lambda}$.  

Next the trace class properties
of the zonal Coulomb and fluctuation operators will be scrutinized.
The corresponding statements will be demonstrated here just 
on $\mathcal H^{(0)}$ by explicit eigenvalue computations.
By formulas 
\begin{equation}
\label{cou_eigval}
{\int_0^\infty r^{2m}e^{-\lambda r^2}dr\over
\int_0^\infty r^{2m+1}e^{-\lambda r^2}dr}={2m-1\over 2m}
{\int_0^\infty r^{2m-2}e^{-\lambda r^2}dr\over
\int_0^\infty r^{2m-1}e^{-\lambda r^2}dr},
\end{equation}
the eigenvalues of $V^{(0,0)}$ regarding the eigenfunctions 
$z^me^{-{1\over 2}\lambda r^2}$ are 
\begin{equation}
Q\sqrt{\pi\lambda}
{1\cdot 3\dots (2m-1)\over   
2\cdot 4\dots 2m }=  
Q\sqrt{\pi\lambda}
{(2m)!\over 2^{2m}(m!)^2}\approx
{Q\sqrt{\pi\lambda}\over \sqrt{\pi m}},
\end{equation}
where the estimation is computed by the Stirling formula 
$n!\approx\sqrt{2\pi}n^ne^{-n}\sqrt n$.
A better approximation,
$Q\sqrt\lambda(4m+1/3)^{1\over 2}(2m+1/3)^{-1}$,
approximating the eigenvalue 
$Q\sqrt{\pi\lambda}$ defined for $m=0$ by the 
finite value $Q\sqrt{3\lambda}$,
can be established by
$n!\approx\sqrt{(2n+1/3)\pi}n^ne^{-n}$. Instead of $0$, the latter 
one approximates $0!=1$ by $\sqrt{\pi/3} \approx 1.02333$.

Therefore, the $V^{(0,0)}$ has infinite trace and 
$L^2$-norm, however, it is in the $L^{2+\epsilon}$-class, for all 
$\epsilon >0$. This statement is true for all zonal Coulomb operators 
and fluctuation operator $F^{(a\to b\to a)}$ satisfying $a\not =b$.
\medskip

\noindent{\bf Remark.} It is a natural question if, instead of the 
3D-potential, the 2D-Coulomb potential should be considered in the zonal
theory. The negative answer becomes clear after 
computing the eigenvalues (\ref{cou_eigval}) corresponding to the 
2D Coulomb potential 
$Q\ln r$. 
Then, the recursion formula for computing the numerator is: 

\begin{eqnarray}
\label{2Dcou_eigval}
\int_0^\infty r^{2m}\ln (r)r e^{-r^2}dr=\\
{1\over 2}\big(r^{2m}\ln (r)e^{-r^2}_{/r=0}
+{2m}\int_0^\infty r^{2m-2}\ln (r)re^{-r^2}dr
+\int_0^\infty r^{2m-1}e^{-r^2}dr\big)
\nonumber\\
={1\over 2}\big(\infty 
+{2m}\int_0^\infty r^{2m-2}\ln (r)re^{-r^2}dr
+2\cdot 4\dots (2m-2)\big),
\nonumber
\end{eqnarray}
according to which the corresponding ``eigenvalue'' is 
${1\over 2}(\infty +\dots +\infty +{1\over 2m}+{1\over 2(2m-2)}+\dots )
=\infty +A_m$, 
where $A_m\to A>0$.

The infinities and relation $A>0$ mean that the 2D Coulomb operator
does not define appropriate zonal transmission and fluctuation operators. 
Actually, the 2D Coulomb operator does not properly describe the physical
situation considered in this paper. 
In fact, this operator assumes a flat
2-dimensional nucleus, whereas, in the 
zonal theory, the  
charged particle is only orbiting in a 2D plane about the 
nucleus which provides radiation field according
to the 3D Coulomb law.
This physical situation is described by 
means of the complex Heisenberg group representation
in terms of the canonical 
coordinates defined on this plane. Also note that the 2D Landau operator 
is defined by a 3D Zeeman operator such that one drops only
the 1-dimensional Laplace operator $\partial^2_{z}$ from the 3D
operator. This means that the particle has zero z-kinetic
energy, i. e., it is not moving into the z-direction. In other words, only
the movements of the 3D particles are restricted onto planes, but they
are not considered to be 2D objects.
Thus really the 3D-potential
is right to consider in the zonal theory. 
In higher dimensions the right
Coulomb potential is 
$Q^\kappa/r^\kappa$. 
\subsection{Lamb shift}

The Lamb shift, experimentally measured by Willis Lamb and Robert
Retherford in 1947, is a small difference in energy between two energy 
levels $2s_{1 / 2}$ and $2p_{1 / 2}$ of the hydrogen atom in quantum 
mechanics. According to Dirac and Schr\"odinger theory, 
hydrogen states with the same $n$ and $j$ quantum numbers but 
different $l$ quantum numbers ought to be degenerate.
However, the experiment pointed out that this was not so - that the 
$2p_{1/2}(n=2,l=1,j=1/2)$ state is slightly lower than the 
$2s_{1/2}(n=2,l=0,j=1/2)$ state resulting 
in a slight shift, $\approx 1000 MHz$, of the corresponding spectral 
line (the Lamb shift).
It might seem that such a tiny effect would be deemed insignificant, 
but in this case that shift probed the depths of our understanding of 
electromagnetic theory.
It became the major stimuli for renewed interest for finding effective 
tools which can deal with the infinities invading QED.
This renewed struggle resulted the renormalization technique which is
the only one, even today, in overcoming this enormous difficulty.

It was long suspected that a possible explanation might be the shift
of energy levels by the interaction of the electron with the radiation
field. This shift comes out infinite in all existing theories.
In 1947, Hans Bethe \cite{be,s.j.} was the first to explain 
the Lamb-shift in the 
hydrogen spectrum. He followed the general 
ideas of Kramers on mass renormalization. The actual calculations are
non-relativistic, whose short description is as follows. 

Due to its interaction with transverse electromagnetic waves, 
the self-energy of an electron in a quantum state $m$ is \cite{h}:
\begin{eqnarray}
\label{w}
W=
-{2e^2\over 3\pi\hbar c^3}
\int_0^\infty k\sum_n{|\mathbf v_{mn}|^2
\over E_n-E_m+k}dk
=\\
-{2e^2\over 3\pi\hbar c^3}
\int_0^\infty\big( 
\sum_n|\mathbf v_{mn}|^2-
\sum_n{|\mathbf v_{mn}|^2(E_n-E_m)
\over E_n-E_m+k}\big)dk,
\end{eqnarray}
where $k=\hbar\omega$ is the energy of the quantum and 
$\mathbf v=\mathbf p/m_e=(\hbar/\mathbf im_e)\nabla$ 
is the velocity in the non-relativistic theory.
The second line decomposes into the sum, $W_0+W^\prime$, of two
divergent integrals, where $W_0$
represents the change of the kinetic energy of the electron for fixed
momentum, due to the fact that electromagnetic mass is added to the
mass of the electron. According to the mass renormalization of Kramers,
this energy should be disregarded, because this electromagnetic mass is
already contained in the experimental electron mass. Therefore the
relevant part of the self-energy becomes $W^\prime =W-W_0$, which is 
considered to be the true shift of the levels due to radiation 
interaction. 
Integral defining $W^\prime$ is still logarithmically divergent which is
renormalized by cutting off the high frequencies $k$ at  
$K\approx m_ec^2$. Bethe assumed that a relativistic hole-theoretic
calculation would provide an explanation for this natural cutoff.
After the third assumption, which considers $\ln (K/|E_n-E_m|)$
as a constant 
(independent of $n$) number, these calculations provided 
$W_{2s}^\prime\approx 1040MHz$, which was in excellent agreement with 
the observed value of $\approx 1000MHz$.

The formulas describing the Lamb shift in a general situation are
$
\Delta_{Lamb}=\alpha^5m_ec^2{k(n,0)\over4n^3},
$
for $l=0$ with $k(n,0)$ around 13 varying slightly with $n$, and
\begin{equation}
\label{lamb}
\Delta_{Lamb}=\alpha^5m_ec^2{1\over4n^3}\big (k(n,l)\pm
{1\over \pi(j+{1\over 2})(l+{1\over 2})}\big),
\end{equation}
for $l\not =0$ and $j=l\pm{1\over 2}$ (inner quantum number, introduced
by Sommerfeld),
with $k(n,l)$ a small number $(< 0.05)$, furthermore,
$\alpha=e^2/\hbar c4\pi\epsilon_0\approx 7.297352568(24)
\times 10^{-3}\approx 1/137$ denotes the fine structure constant.

\subsection{Bethe's computation in light of the zonal theory.}

On a zone the complete physics is determined by the complex Heisenberg
group representation (\ref{rho}). On the Fock zone, where the
computations are carried out, the magnetic quantum number $m$ is equal
to the azimuthal quantum number $l$ and the normalized eigenfunctions
are
\begin{equation}
\Psi_l={\lambda^{l+1\over 2}\over\sqrt{l!\pi}}z^l
e^{-{1\over 2}\lambda r^2}
=\psi_l
e^{-{1\over 2}\lambda r^2}.
\end{equation} 
Thus for the velocity we have:
\begin{equation}
\mathbf v_{m}=(\partial_z\psi_l)
e^{-{1\over 2}\lambda r^2}=\sqrt{l\lambda}\Psi_{l-1},
\end{equation}
therefore,
$|\mathbf v_m|^2=l\lambda =m\lambda$. Note that the only non-trivial
components of
$\mathbf v_{mn}$ resp. $E_n-E_m$
are
$\mathbf v_{m(m-1)}$ resp. $E_{m-1}-E_m=E_m/(2m-1)$, thus Bethe's
third assumption of $E_n-E_m$ being independent of $n$ is 
automatically satisfied. Thus $\sum_n$ in (\ref{w}) consists only of 
one term corresponding to $n=m-1$. 

Next the implications caused by choosing the cutoff constant 
$K\approx m_ec^2$ on the zonal setting is explained. In the Bohr model 
of the hydrogen atom the Rydberg constant
$
Ry=h c R_\infty = \frac{m_e c^2 \alpha^2}{2},   
$
where
$
R_\infty = \frac{\alpha^2 m_e c}{4 \pi \hbar},
$
is defined as the energy on the innermost
energy level. This level has index $1$ and the electron 
is not resting there. The resting state corresponding to the index $0$ 
is not considered in that theory because it leads to false
conclusions. The energy is infinity there, anyway. 
In our zonal model, however, the innermost
energy level on $\mathcal H^{(a)}$ is represented by the eigenfunction 
$\overline z^a$ for which the velocity is zero: $\mathbf v_0^{(a)}=0$.
(This condition does not contradicts uncertainty principle because 
the zonal particles can not be localized at one point.) 
Therefore, it is reasonable to choose the Rydberg constant $R_y^{(a)}$ 
in the zonal theory to be the rest energy $m_ec^2$. More precisely, 
the spectrum for $V^{(a,a)}$ is defined such that
the dimensionless spectrum, defined by assuming $Q=\lambda =1$, is 
multiplied with $m_ec^2$.
Since the discrete spectrum of 
the zonal Coulomb field is decreasing and the highest dimensionless 
eigenvalue is $1$, the $m_ec^2$ becomes the highest
energy level possible. In short, choosing cutoff constant $K$
means choosing new Rydberg constant on the zonal setting. 
By the definition of $V^{(a,a)}$, the original one is $Q\sqrt \lambda$.
It should be mention yet that the interaction theory developed here
prefers $m_ec^2$ for the zonal Rydberg constant.

The greatest difficulty  is created by 
the measure $dk$, measuring its domain, $[0,\infty )$, by infinity.
The problems persist to exist  
even after introducing the cutoff constant. 
For instant, it still defines infinite energy summations.  
These difficulties seem to be
originated from lacking the probabilistic feature characteristic
for those measures which appear in quantum theory. We attack
this problem right at this point! Bethe was compelled to use a cutoff
constant along with other renormalization techniques because he 
had not the right finite measure allowing
finite summations in store. Thus for the solution of this problem 
one should find the appropriate finite measure defining finite 
summations which directly produces the desired Lamb shift
formula without any further assumption.  In the zonal theory,
where the multiplicity of each eigenvalue of the discrete 
zonal Coulomb spectrum is $1$, one has a new mathematical 
and physical situation which makes it possible to find this measure.
At this point we depart from Bethe's computations and give a new start
to solving this problem.

\subsection{The quantum hypothesis untying the Gordian knot.}

This intricate tangle of difficulties will be untied by
a quantum hypothesis similar to the Planck hypothesis. The historic 
Planck hypothesis concerns the amount of
energy $U(\nu )d\nu$ radiated by the blackbody in the frequency 
range between $\nu$ and $\nu +d\nu$. 
By supposing equal likelihood for radiating $U(\nu )$ (equipartition
principle), the old theory yielded the Rayleigh-Jeans law: 
$U(\nu )d\nu\sim\kappa T\nu^2d\nu$, which
contradicts the empirical curve described by the Wien law: 
$U(\nu )d\nu\sim\nu^3e^{-h\nu/\kappa T}d\nu$
($\kappa$ and $T$ denote the Boltzmann constant and temperature 
respectively).
The controversy arising between theory and experiment was resolved by 
Planck by the hypothesis that the energy
attached to frequency $\nu$ is restricted to the integral multiple
of the basic unit $h\nu$, i. e. $E_n=nh\nu$, where $n$ is
any positive integer number. Furthermore, the preprobability that the 
wall emits-absorbs an energy-quanta $E_n$ is 
$\tilde W(n)\sim e^{-E_n/\kappa T}=e^{-nh\nu/\kappa T}$. 
Thus, by normalization, the probability is
$W(n)=e^{-nh\nu/\kappa T}(1-e^{-nh\nu/\kappa T})$.
This hypothesis yields the Wien law.

In the present situation
the quantization of the energy due to the interaction of a Landau  
electron with the zonal Coulomb field should be established by an 
adequate preprobability amplitude (finite complex measure). 
It is rather apparent that in
integral formula (\ref{w}) the electron of magnetic
quantum state $m$ interacts
with the radiation field of energy level $k$ according to
the equipartition principle and not by a preprobabilistic amplitude
which gives high priority for certain energies $k$ while small
chances for the other ones. By applying cutoff constants and other
regularization-s, the renormalization theory works down, actually, 
this equipartition-amplitude to a preprobabilistic one. 
In our approach we write up the adequate interaction preprobability
amplitude at the very beginning whose adequacy will be probed
by testing if it really provides the right Lamb shift formula.

This interaction amplitude is written up for fixed
dimensionless energy levels  
\begin{equation}
\epsilon_p=(4p+2)\lambda\quad \textrm{and}\quad
\epsilon_B=4\lambda^2,\quad \textrm{where}
\quad\lambda =1,
\end{equation} 
of the free electron resp. constant magnetic field and the 
dimensionless energy levels, $E_k^{(a)}$, of the zonal Coulomb field 
defined on a zone $\mathcal H^{(a)}$. In this review we give explicit 
computations only on the Fock zone ($a=0$) where the dimensionless 
Coulomb eigenvalues are 
\begin{equation}
\label{e_k}
E_k^{(0)}=\sqrt{\pi}
{(2k)!\over 2^{2k}(k!)^2}=\sqrt{\pi}
{\Gamma (2k+1)\over 2^{2k}\Gamma^2(k+1)}=\sqrt\pi G_k^{(0)}.
\end{equation}
In terms of the Gamma function, such explicit formulas can be 
established also for the other zones. These formulas define 
dimensionless energies
not just for the discrete values $k=0,1,\dots $ but also for
real numbers $k\geq 0$, allowing not just
discrete but also continuous interaction amplitudes. Actually the
discrete amplitude can be defined by means of the continuous one
and the summations in the discrete case can be approximated
by the corresponding integral defined for the continuous version.
The exact form of the hypothesis prepared by these remarks is as follows.

\newpage
      
{\bf Hypothesis for the preprobabilistic amplitude (finite complex
measure) of the 
electromagnetic interaction.} 
{\it The energy increment  
for an electron which is in the dimensionless energy state 
$\epsilon_p$ resp. $\epsilon_B$ 
and interacts
with a dimensionless zonal radiation field $V^{(a,a)}$ can be 
determined by the preprobabilistic amplitude (finite complex measure)
\begin{eqnarray}
\label{amp}
\mathcal A^{(a)}_{\mathbf i}(\epsilon_.,k)dk=
\sqrt{m_ec^2}\alpha^{5\over 2}{1\over 2} 
de^{-{1\over 2}\sqrt\pi\epsilon_. E^{(a)}_k\mathbf i}\\
=\sqrt{m_ec^2}\alpha^{5\over 2}{1\over 2} 
({-{\pi\over 2}\epsilon_. G_k^{(a)\prime}\mathbf i})
e^{-{1\over 2}\pi\epsilon_. G^{(a)}_k\mathbf i}dk
\nonumber\\
=\sqrt{m_ec^2}\mathcal B^{(a)}_{\mathbf i}(\epsilon_.,k)dk,
\label{b_i}
\end{eqnarray}
where $\epsilon_.$ stands for 
$\epsilon_p$ or $\epsilon_B$, such that
the amplitude for the energy shift,
due to the interaction with the zonal Coulomb field, is the
finite value $\Sigma^{(a)}_{\mathbf i} (\epsilon_.)=\int_{k =0}^{\infty}
E^{(a)}_k \mathcal A^{(a)}_{\mathbf i}(\epsilon_.,k)dk$ and, therefore, 
the corresponding Lamb shift
(energy increment) is 
$\Delta^{(a)}_{Lamb}=\Sigma^{(a)}_{\mathbf i} (\epsilon_.)
\overline\Sigma^{(a)}_{\mathbf i} (\epsilon_.)$.

The discrete density $\mathcal D_{\mathbf i}(\epsilon_.,k)$ is 
defined by 
restricting the continuous density
$\mathcal A^{(a)}_{\mathbf i}(\epsilon_.,k)$
to the discrete set 
$k=0,1,2,\dots $ of numbers. Then the summations corresponding 
to the above
integrals can be approximated by the above integrals.

The scalar density, 
$\mathcal A^{(a)}_{1}(\epsilon_.,k)dk$, is defined by omitting 
$\mathbf i$ from
the above formulas. This density can be used for further studying
Bethe's computations (these details are omitted). 
} 

In the density formula (\ref{amp}) all quantities are dimensionless
except $m_ec^2$ which is the energy of the resting electron.
The fine-structure constant $\alpha\approx 1/137$ is the fundamental 
physical constant characterizing the strength of the electromagnetic 
interaction. It is a dimensionless quantity, and thus its numerical 
value is independent of the system of units used.
It can be thought of as the square of the ratio of the elementary 
charge to the Planck charge.
For any arbitrary length $s$ used in formulas
$
 2 \pi s = \lambda = \frac{c}{\nu} 
$ 
of classical quantum theory, the $\alpha$ is the ratio of two energies.
Its value cannot be predicted by the theory, and has to be inserted 
based on experimental results. In fact, it is one of the twenty-odd 
"external parameters" in the Standard Model of particle physics.
This also means that the above Hypothesis is independent from the other
ones of quantum theory which has to be objected to experimental
testing. Below we show that this Hypothesis implies the Lamb shift
formula, thus this testing is already done by the Lamb-Retherford
experiment. 
Thus the positively tested Hypothesis lifts out $\alpha$
from the set of the twenty-odd external parameters and turns it into 
an internal, equal partner to the Planck constant.

At this point of argumentation it is appropriate to quote from the 
preface written to the ``Pauli Lectures on Physics, Volume 1." 
by the last assistant of Pauli:
``For Pauli the central problem of electrodynamics was the field
concept and the existence of an elementary charge which is expressible
by the fine structure constant $e^2/mc=1/137$. This fundamental pure 
number had greatly fascinated Pauli, as can be seen from the list of 
references to his work assembled in the appendix. For Pauli the 
explanation of the number $137$ was the test of a successful field
theory, a test which no theory has passed up to now. This number $137$
transcendented into a magic symbol at Pauli's death. When I visited
Pauli in the hospital, he asked me with concern whether I had noticed
his room number: 137! It is in this room that he died a few days later.
Charles P. Enz, Geneva. 17 November 1971."

This quote inspires the following reformulation of the above ideas 
developed for the complex plane $\mathbf C$.

{\bf Reformulated Hypothesis in terms of Dimensionless
Quantum Theory ($D_{less}QTh$).} {\it The dimensionless quantum
theory is based on the complex Heisenberg algebra representation
(\ref{rho}) defined for $\lambda =1$. 
The irreducible subspaces, 
$\mathcal H^{(a)}$, 
of this reducible representation are the 
so called dimensionless Zeeman zones among which 
$\mathcal H^{(0)}$ is the Fock space. The dimensionless Hamiltonian 
associated with this representation is  
\begin{equation}
\label{DlessBox}
-\Box
=-(\Delta_X - 2\mathbf i D\bullet -4(1 + 
\frac 1 {4} |X|^2)),
\end{equation}
where $D\bullet$ is defined by the complex structure $J$ involved
to mathematical modeling. This operator is the sum of the dimensionless
Landau operator and a constant Hamiltonian belonging to the constant
magnetic field. The dimensionless spectrum of $\Box$ is
$
-((4p+2) +4),  
$
where $p=n+l$ is established earlier as a quantum number for the 
Landau electron and $4$ belongs to the constant magnetic field.
Each eigenvalue in this spectrum has infinite multiplicity and the
the zones are established by sorting out the eigenfunctions such that,
on a zone, each multiplicity becomes $1$ and any two zones are 
isospectral.

The dimensionless Coulomb potential is $1/r$ which defines a 
dimensionless Coulomb integral operator $V^{(a,a)}$ for each zone.
This zonal Coulomb operator commutes with the Landau operator and
operator $-\Box^{(a)}-V^{(a,a)}$ is the complete dimensionless
zonal Zeeman Hamiltonian. The dimensionless Coulomb eigenvalues
on the Fock zone are described by (\ref{e_k}). In terms of Gamma 
function and index $(a)$, they can be explicitly computed for all zones.

The dimensionless elementary charge is defined by 
$\aleph =\big({1\over 4}\alpha^5\big)^{1\over 6}\approx {1\over 76}$ 
(see the explanation below) and the
dimensionless preprobabilistic amplitude 
$\mathcal B^{(a)}_{\mathbf i}(\epsilon_.,k)dk$ is
defined 
in terms of the cubed elementary charge 
${1\over 2}\alpha^{5\over 2}=\aleph^3$ and an elementary density 
in formula (\ref{b_i}). 
Then the dimensionless energy 
due to the interaction between
the electron, having the quantum numbers $\epsilon_p=4p+2$ resp. 
$\epsilon_B=4$, and the zonal Coulomb field can be determined as follows.
The dimensionless
amplitude for this energy is
$\beta^{(a)}_{\mathbf i} (\epsilon_.)=\int_{k =0}^{\infty}
E^{(a)}_k \mathcal B^{(a)}_{\mathbf i}(\epsilon_.,k)dk$ and the total
amplitude, 
$\beta^{(a)}_{\mathbf i,total}$,
is the sum of the two amplitudes.
The corresponding dimensionless Lamb shifts
(dimensionless energy increments due to interactions) are 
$\Delta^{(a)}_{D_{less}Lamb}(\epsilon_.)
=\beta^{(a)}_{\mathbf i} (\epsilon_.)
\overline\beta^{(a)}_{\mathbf i} (\epsilon_.)$ and
$\Delta^{(a)}_{D_{less}Lamb}(total)
=\beta^{(a)}_{\mathbf i,total} 
\overline\beta^{(a)}_{\mathbf i,total}$,
where the last formula is based on the independence of the Hamiltonian
of the constant magnetic field from the rest part of the Hamiltonian.
(This independence is exhibited by the commutativity of 
these operators.)
For an electron in quantum state $\epsilon_.=\epsilon_p$, or, 
$\epsilon_B$, the true Lamb shift having complete physical dimensions is
\begin{eqnarray}
\Delta^{(a)}_{Lamb}(\epsilon_.)=m_ec^2
\Delta^{(a)}_{D_{less}Lamb}
(\epsilon_.),\\
\Delta^{(a)}_{Lamb}(total)=m_ec^2
\Delta^{(a)}_{D_{less}Lamb}(total).
\end{eqnarray}

Instead of one electron one can consider also a zonal charge-field, 
defined by the density $\rho^{(a)}_e(z)$, such that at each point 
$z\in\mathbf C$ the charge is in the total quantum state determined by  
$\epsilon_p$ resp. $\epsilon_B$. Then the energy due to the  
interaction of the charge-field with the zonal Coulomb field is
\begin{equation}
\Delta^{(a)}_{Lamb}(total)=
\big(\int \rho^{(a)}_ec^2dz\big)
\Delta^{(a)}_{D_{less}Lamb}(total).
\end{equation}

}
In Bethe's formula, obtained for the Lamb shift, constant
${1\over 4}\alpha^5m_ec^2$ appears in the form ${1\over 2}\alpha^3Ry$,
where $Ry=hc/\lambda =hc/(c\nu)=h\nu$ is the energy on the innermost
energy level of the hydrogen atom. This constant includes $\alpha$ by
the following computations:
$$
    R_\infty = \frac{\alpha^2 m_e c}{4 \pi \hbar},\quad
Ry=h c R_\infty = \frac{m_e c^2 \alpha^2}{2}.   
$$
The Rydberg constant is defined by means of the Bohr model of hydrogen 
atom where the innermost level has index $1$ and the electron is not
resting there. Considering the resting state with index $0$ leads to 
false conclusions in that model. In our zonal model, however, the 
innermost energy level on $\mathcal H^{(a)}$ is 
represented by the eigenfunction 
$\overline z^a$ for which the velocity is zero: $\mathbf v_0^{(a)}=0$. 
Therefore, in the zonal theory, the Rydberg constant $R_y^{(a)}$ can be
defined by the rest energy $m_ec^2$. If one considers Bethe's formula
as a natural pattern and assumes that the Lamb shift
constant should appear in terms of an 
elementary charge $\aleph$ 
in the form $\aleph^6R_y^{(a)}$  
then this elementary charge must satisfy $\aleph^6={1\over 4}\alpha^5$.
In this interpretation both the electron and the nucleus are supposed
to be charged with the elementary charge.
This is the major motivation for defining the elementary charge in the
form described in the Hypothesis.
\subsection{Testing the Hypothesis by the Stirling approximations.}

The real test of the Hypothesis is if it really provides the
Lamb shift formula (\ref{lamb}). 
The computations are demonstrated below on the Fock zone and,
instead of (\ref{amp}), an 
approximating density defined by the Stirling approximation of $n!$
will be used for computations.
To avoid divergence at $0$, the second Stirling formula is the 
appropriate one for defining 
$
E^{(0)}_k\approx (4k +1/\pi )^{1\over 2}/(2k+1/\pi )=S_k
$. 
As opposed to the original version, where $3$ is used instead of $\pi$, 
this Stirling formula approximates $0!$ by $1$.

Thus the approximating Stirling density for 
$\epsilon_p=4p+2=4l+2=\epsilon_l$ is
\begin{eqnarray}
de^{-\sqrt\pi (2l+1)S_k\mathbf i}=
{-\sqrt\pi (2l+1)S_k^\prime\mathbf i}
e^{-\sqrt\pi (2l+1)S_k\mathbf i}dk
\end{eqnarray}
and the amplitude for the shift of the self energy of the electron 
due to the interaction with the zonal Coulomb field can be found by
the following computations:
\begin{eqnarray}
\sigma_{\mathbf il}^{(0)}=
\int_{0}^\infty S_kde^{-\sqrt\pi (2l+1)S_k\mathbf i}=
\int_{0}^\infty S_k
\big(e^{-\sqrt\pi (2l+1)S_k\mathbf i}\big)^\prime dk=
\end{eqnarray}
\begin{eqnarray}
=-\sqrt\pi
e^{-(2l+1)\pi\mathbf i}
-\int_{k=0}^\infty S_k^\prime 
e^{-\sqrt\pi (2l+1)S_k\mathbf i}dk=
\nonumber
\end{eqnarray}
\begin{eqnarray}
=\sqrt \pi
+{1\over\sqrt\pi (2l+1)\mathbf i}
\int_0^\infty\big(e^{-\sqrt\pi (2l+1) S_k\mathbf i}\big)^\prime dk=
\nonumber
\end{eqnarray}
\begin{eqnarray}
=\sqrt \pi
+{1\over\sqrt\pi (2l+1)\mathbf i}(1-
e^{-(2l+1)\pi\mathbf i})
=\sqrt \pi -
{\mathbf i\over\sqrt\pi (l+{1\over 2})},
\nonumber
\end{eqnarray}
\begin{eqnarray}
\Delta^{(0)}_{Lamb}(\epsilon_p)=
m_ec^2\alpha^5{1\over 4}
\sigma^{(0)}_{\mathbf il} 
\overline\sigma^{(0)}_{\mathbf il} 
=m_ec^2\alpha^5{1\over 4}
(\pi
+{1\over\pi (l+{1\over 2})^2})
\end{eqnarray}

The Hamilton operator of the constant magnetic field commutes with the
complete operator, thus the complete amplitude is the sum of 
$\sigma^{(0)}_{\mathbf il}$ and
$\sigma^{(0)}_{\mathbf iB}$.
The above computation provides the latter amplitude by substituting
$(2l+1)$ by $2$. Then we have
\begin{equation}
\sigma^{(0)}_{\mathbf iB}=-\sqrt\pi,\quad
\sigma_{total}^{(0)}=\sigma^{(0)}_{\mathbf il}+
\sigma^{(0)}_{\mathbf iB}=-
{\mathbf i\over\sqrt\pi (l+{1\over 2})},
\end{equation}
\begin{equation}
\Delta^{(0)}_{total}=
{m_ec^2\alpha^5\over 4\pi (l+{1\over 2})^2}.
\end{equation}
Although term $\pi^{-1}(l+{1\over 2})^{-2}$ can be written in form
\[
{1\over\pi(l+{1\over 2})^2}=
{1\over 2\pi(l+1)(l+{1\over 2})^2}
+{1\over\pi(l+1)(l+{1\over 2})},
\]
where the first term on the right side is a decreasing sequence 
determining small numbers
$<\,0.03536$ for $l\geq 1$, this first term can not be identified
with $k(n,l)$ of the experimental formula (\ref{lamb}). The $k(n,l)$
is due to other interactions not considered in this paper. The formula
established here concerns just the interaction with the Coulomb field,
which is the main contribution to the Lamb shift. Even this main term 
is established in (\ref{lamb}) by considering the inner quantum number 
$j$. This quantum number has net appeared on the scene yet and will 
furnished into the zonal theory after introducing the
spin concept. This is why 
$(l+{1\over 2})^2$ appears in the place of 
${1\over\pi(l+{1\over 2})(j+{1\over 2})}$
in the scalar theory presented in this paper. The discrepancies 
regarding the radial quantum number $n$ are due to the discrepancies
between the classical and zonal radial quantum numbers.
The zonal one is included into $p=n+l$ and it is zero on the Fock zone.
Thus $1/n^3$ in formula (\ref{lamb}) does not make any sense there.
Furthermore, the dependence on $n$ seems to be just quadratic on the 
zonal setting. However, the explicit formulas of Lamb shift on the
higher order zones reveal a higher order dependence on the radial
quantum number. 
  
\medskip
\noindent{\bf Acknowledgements.} The author is indebted for the hospitality
and excellent working conditions provided by the Max Planck Institute 
for Mathematics in the Sciences, Leipzig, in the academic year 2007/2008. 
My particular gratitude yields to Prof. Eberhard Zeidler for the
conversations about several parts of this paper.


\begin{thebibliography}{9}
\bibitem[AB]{ab}
Y. Aharanov and D. Bohm:
\newblock Significance of electromagnetic potentials in the
quantum theory.
\newblock {\em Phys. Rev.}, 115:485--491, 1959.

\bibitem[AC]{ac}
Y. Aharanov and A. Casher:
\newblock Ground state of spin-$1/2$ charged particle in a
two-dimensional magnetic field.
\newblock{\em Phys. Rev. A},
19:2461-2462, 1979.


\bibitem[B]{b}
V. Bargmann:
\newblock On a Hilbert space of analytic functions and an 
associated integral
transform. Part I.
\newblock{\em Comm. Pure Appl. Math.}, 14:187-214, 1961

\bibitem[Be]{be}
H. A. Bethe:
\newblock The electromagnetic shift of energy levels.
\newblock{\em Physical Review}, Vol. 72, p. 339, 1947 

\bibitem[Foc]{foc}
V. Fock:
\newblock Konfigurationsraum und zweite Quantelung.
\newblock{\em Zeit. f\"ur Fys.} 75:622-647, 1932.

\bibitem[H]{h}
W. Heitler:
\newblock The quantum theory of radiation.
\newblock{\em Dover Publications}, 1984.

\bibitem[I]{i}
K. It\^o:
\newblock Complex multiple Wiener integral. 
\newblock{\em Jap. J. Math.} 22:63--86 (1953). 

\bibitem[K]{k}
A. Kaplan:
\newblock Riemannian nilmanifolds attached to Clifford modules.
\newblock{\em Geom. Dedicata}, 11:127--136, 1981.

\bibitem[LL]{ll}
L. D. Landau, E. M. Lifshitz:
\newblock Quantum Mechanics.
\newblock Pergamon Press LTD, 1958.

\bibitem[P]{p}
W. Pauli:
\newblock Pauli Lectures on Physics.
\newblock Dover Publications, 2000.

\bibitem[S.J.]{s.j.}
J. Schwinger (editor):
\newblock Selected papers on QED.
\newblock Dover Publ., 1958.

\bibitem[S.S.]{s.s.}
S. S. Schweber:
\newblock QED and the men who made it: Dyson, Feynman, Schwinger, and
Tomonaga.
\newblock Princeton Univ. Press, 1994.

\bibitem[Sh]{sh}
I. Shigekawa: 
\newblock Itô-Wiener expansions of holomorphic functions on the 
complex Wiener space. 
\newblock {\em Stochastic analysis}, 459--473, 
Academic Press, Boston, MA, 1991. 

\bibitem[Su]{su}
H. Sugita: 
\newblock Holomorphic Wiener function.  
\newblock New trends in stochastic analysis (Charingworth, 1994), 
399--415, World Sci. Publishing, River Edge, NJ, 1997. 

\bibitem[Sz1]{sz1}
Z. I. Szab\'o:
\newblock Locally non-isometric yet super isospectral spaces.
\newblock{\em Geom. funct. anal. (GAFA)}, 9:185--214, 1999.

\bibitem[Sz2]{sz2}
Z. I. Szab\'o:
\newblock Isospectral pairs of metrics on 
balls, spheres, and other manifolds with different local geometries.
\newblock{\em Ann. of Math.}, 154:437--475, 2001.

\bibitem[Sz3]{sz3}
Z. I. Szab\'o:
\newblock A cornucopia of isospectral pairs of metrics on 
spheres with different local geometries.
\newblock{\em Ann. of Math.}, 161:343--395, 2005.

\bibitem[Sz4]{sz4}
Z. I. Szab\'o:
\newblock Reconstructing the intertwining operator and new striking
examples added to ``Isospectral pairs of metrics on balls and 
spheres with different local geometries".
\newblock{\em DG/0510202 (submitted)}

\bibitem[Sz5]{sz5}
Z. I. Szab\'o:
\newblock Theory of zones on Zeeman manifolds: 
A new approach to the infinities
of QED.
\newblock {\em DG/0510660 (submitted)}

\bibitem[Sz6]{sz6}
Z. I. Szab\'o:
\newblock Normal zones on Zeeman manifolds with trace class heat and 
Feynman kernels and well defined zonal Feynman integrals.
\newblock {\em math.SP/0602445 (submitted)}

\bibitem[Sz7]{sz7}
Z. I. Szab\'o:
\newblock Pauli-Dirac operators and anomalous zones on Zeeman manifolds.
\newblock submitted.

\bibitem[T1]{t1}
A. Tonomura, et al: 
%N. T. Matsuda, R. Suzuki, A. Fukuhara, N. Osakabe,
%H. Umezaki, J. Endo, K. Shinagawa, Y. Sugita, and H. Fujiwara:
\newblock Observation of Aharanov-Bohm effect by electron holography.
\newblock {\em Phys. Rev. Lett.}, 48:1443--1446, 1982.

\bibitem[T2]{t2}
A. Tonomura, et al: 
%N. Osakabe, T. Matsuba, T. Kawasaki, J. Endo, S. Yano, and H. Yamada:
\newblock Evidence for Aharanov-Bohm effect with magnetic field
completely shielded from electron wave.
\newblock {\em Phys. Rev. Lett.}, 56:792-795, 1986.

\bibitem[Z]{s.s.}
E. Zeidler:
\newblock Quantum Field Theory I: Basic in Mathematics and Physics.
\newblock Springer-Verlag, 2006.







\end{thebibliography}
\end{document}